\documentclass[showpacs,twocolumn,twoside,amsmath,amssymb,pra,superscriptaddress,longbibliography]{revtex4-1}
\newcommand*{\balancecolsandclearpage}{%
  \close@column@grid
  \clearpage
  \twocolumngrid
}
\usepackage{times}
\usepackage{graphicx}
\usepackage{amssymb}
\usepackage{url}
\usepackage{epstopdf}
\usepackage[unicode]{hyperref}
\hypersetup{
   unicode=true,          % non-Latin characters in AcrobatÕs bookmarks
   plainpages=false,
   colorlinks=true,       % false: boxed links; true: colored links
   citecolor=blue,        % color of links to bibliography
}
\begin{document}

\title{Interaction blockade for bosons in an asymmetric double well}

\author{Jayson G. Cosme}

\affiliation{Dodd-Walls Centre for Photonics and Quantum Technology, New Zealand Institute for Advanced Study, Centre for Theoretical Chemistry and Physics, Massey University, Auckland, New Zealand}

\author{Mikkel F. Andersen}

\affiliation{Dodd-Walls Centre for Photonic and Quantum Technologies, Department of Physics, University of Otago, Dunedin, New Zealand}

\author{Joachim Brand}

\affiliation{Dodd-Walls Centre for Photonics and Quantum Technology, New Zealand Institute for Advanced Study, Centre for Theoretical Chemistry and Physics, Massey University, Auckland, New Zealand}

\date{\today}                                           % Activate to display a given date or no date

\begin{abstract}
 
The interaction blockade phenomenon isolates the motion of a single quantum particle within a multi-particle system, in particular for coherent oscillations in and out of a region affected by the blockade mechanism. 
For identical quantum particles with Bose statistics, the presence of the other particles is still felt by a bosonic stimulation factor $\sqrt{N}$ that speeds up the coherent oscillations, where $N$ is the number of bosons.
Here we propose an experiment to observe this enhancement factor %using an asymmetric double well 
with a small number of bosonic atoms. The proposed protocol realises an asymmetric double well potential with multiple optical tweezer laser beams. The ability to adjust bias independently of the coherent coupling between the wells allows the potential to be loaded with different particle numbers while maintaining the resonance condition needed for coherent oscillations. 
Numerical simulations with up to three bosons in a realistic potential generated by three optical tweezers predict that the relevant avoided level crossing can be probed and the expected bosonic enhancement factor observed.

\end{abstract}

\maketitle

\section{Introduction}

The high controllability of ultracold atoms provide a natural testbed for observing both novel and well-established quantum many-body phenomena \cite{Bloch2008}. Exquisite control in the level of single neutral atoms has been demonstrated in experiments such as single-atom trapping \cite{Andersen2010}, laser cooling to the quantum mechanical ground state \cite{Kaufman2012,Sompet2016}, and demonstration of the Hong-Ou-Mandel effect \cite{Kaufman2014}.
These technological developments in the manipulation of low-entropy quantum states have paved the way for unique opportunities to observe the emergence of many-body effects as the number of atoms is gradually increased \cite{Wenz2013,Jochim2015}. 

The strength of the interaction between particles often dictates the nonequilibrium properties as a consequence of its effect on the available quantum states in a many-body system. Such an example for electrons in quantum dots is the Coulomb blockade phenomenon wherein an applied gate voltage enables exactly one electron to tunnel into the dot \cite{Lambe1969,Averin1986,Meirav1989,Reimann2002}. For neutral atoms, a similar phenomenon can be seen in the so-called Rydberg blockade where only a single atom out of many is excited  within a  blockade radius due to  strong dipole-dipole interaction between atoms in Rydberg states 
\cite{Lukin2001,Johnson2008,Urban2009,Saffman2010,Browaeys2016}. 
When a similar blockade phenomenon is achieved by short-range interactions between neutral atoms, e.g.\ for tunneling dynamics in a double-well potential, it is known as \emph{interaction blockade} \cite{Micheli2004,Capelle2007,Seaman2007,Dounas2007,Lee2008,Schlagheck2010,Cheinet2008,Preiss2015}. 
It is a direct analog of the Coulomb blockade in electronic transport, except that the blockade is mediated by the  van-der-Waals interactions between neutral atoms instead of the Coulomb force.
A particular example is a double-well potential with a small number of strongly-interacting atoms.
Tilting the double-well can offset intra-well interactions and gives rise to discrete plateaus in the  population of a well corresponding to integer particle numbers \cite{Cheinet2008}. The plateaus originate in the finite energy cost for adding a particle to a well and the need to satisfy a resonance condition to allow particles to coherently move between the wells.
When the resonance condition is met, coherent oscillations of a single particle between the wells are possible, and the oscillation frequency is proportional to the splitting of energy eigenstates at an avoided level crossing.

An interesting question concerns the frequency of coherent oscillations in the blockade regime where only a single particle is allowed to tunnel due to the presence of strong interactions. Simple quantum mechanical arguments imply that the oscillation frequency scales with the square root of the number of particles present if the particles are bosons. Such a bosonic enhancement factor of $\sqrt{N}$ was observed in Rydberg blockade experiments when $N$ excitable atoms were present by measuring the Rabi frequency  \cite{Johnson2008,Urban2009,Saffman2010,Browaeys2016}.
In interaction blockade experiments with double-well potentials, however, the bosonic enhancement factor has not been observed directly so far. Instead, a different particle-number dependence  of the oscillation frequency  was observed  \cite{Preiss2015}. The reason is attributed to the fact that the shape of the double-well potential, in particular the effective tunneling barrier, changes when adjusting the tilt to compensate for the interaction energy brought by different particle numbers \cite{Meinert2013,Preiss2015}. The motivation for the current work is to propose an improved experimental design that  can overcome these limitations and allow for the bosonic enhancement factor to be measured directly from the frequency of coherent oscillations.   

In this work, we study few-body dynamics in an asymmetric double well. We demonstrate by means of numerical simulations the feasibility of observing the expected $\sqrt{N}$ scaling behavior of the oscillation frequencies, or equivalently the energy splittings, in the interaction blockade limit. Specifically, we propose to use an asymmetric double well configuration formed by superimposing multiple optical tweezer potentials. We then simulate the full quantum system for up to $N=3$ bosons in the three-dimensional trapping geometry using the multiconfigurational time-dependent Hartree method for bosons (MCTDHB) \cite{AlonEtAl2008}, obtaining good agreement with a reduced description in terms of an effective Bose-Hubbard dimer model. The paper is organized as follows: Section~\ref{sec:BH} contains a brief introduction of the interaction blockade phenomenon as understood from the Bose-Hubbard model and an analytical prediction from this dimer model for deviations from the expected $\sqrt{N}$ behavior due to finite interaction strengths; Section~\ref{sec:tweezer} describes important considerations in constructing the three-dimensional optical tweezer potentials used in our study; Section~\ref{sec:blockade} presents the main numerical results on the avoided level crossings for $N=2$ and $N=3$ together with the nature of the relevant resonance states and how the energy splittings compare with the bosonic enhancement factor predicted in the interaction blockade regime. This section also contains fidelity calculations and time-dependent results for $N=2$ aimed at validating our suggested protocol; Section~\ref{sec:disc} gives a short summary of our findings, technical details for possible experiments, and an outlook. We have also included an appendix for a short description of the MCTDHB method and a discussion of implemented convergence checks for the numerical approach as well as the determination of effective interaction parameters.

\section{Bose-Hubbard model}\label{sec:BH}

 \begin{figure}[!htb]
\includegraphics[width=0.5\columnwidth]{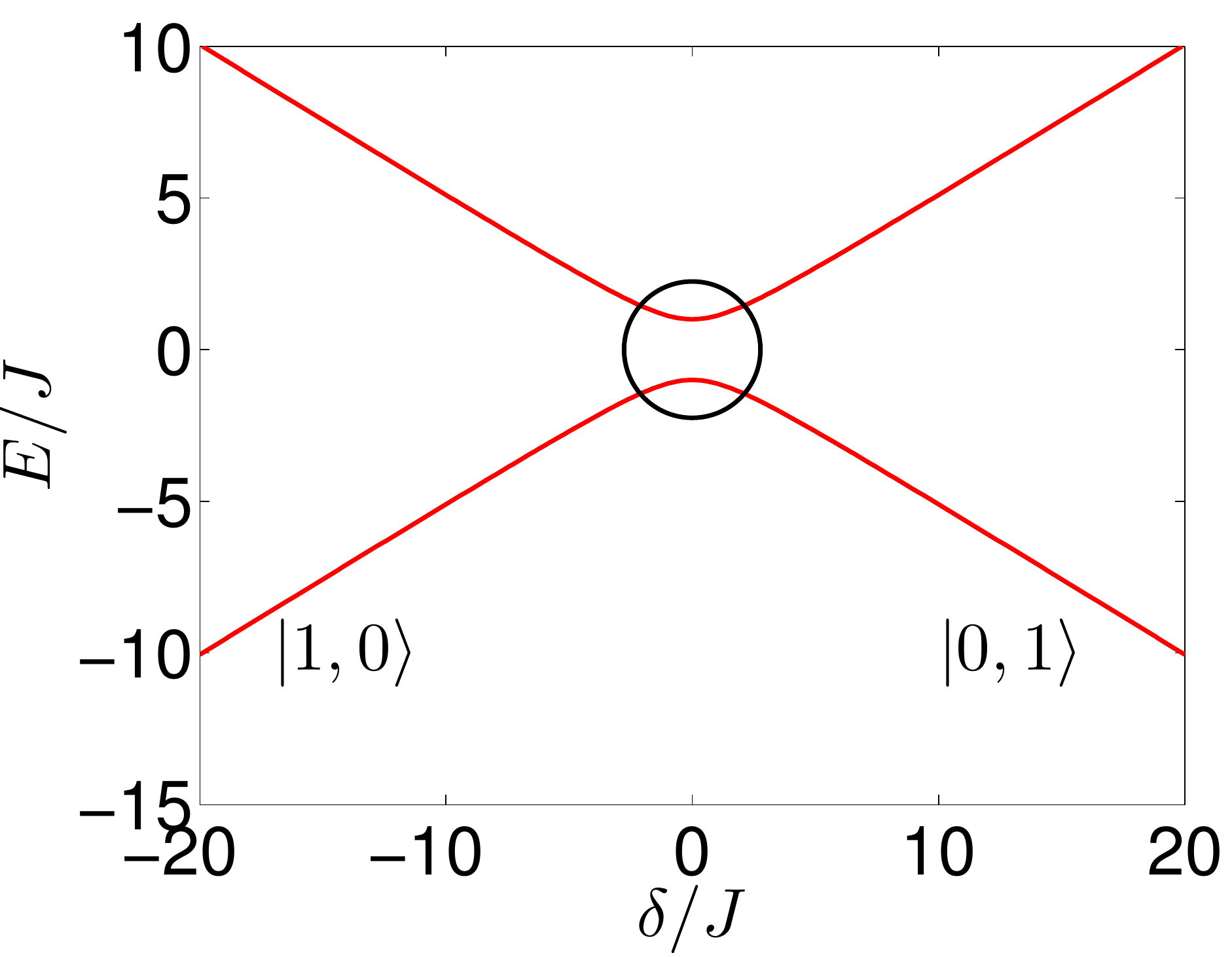}\includegraphics[width=0.5\columnwidth]{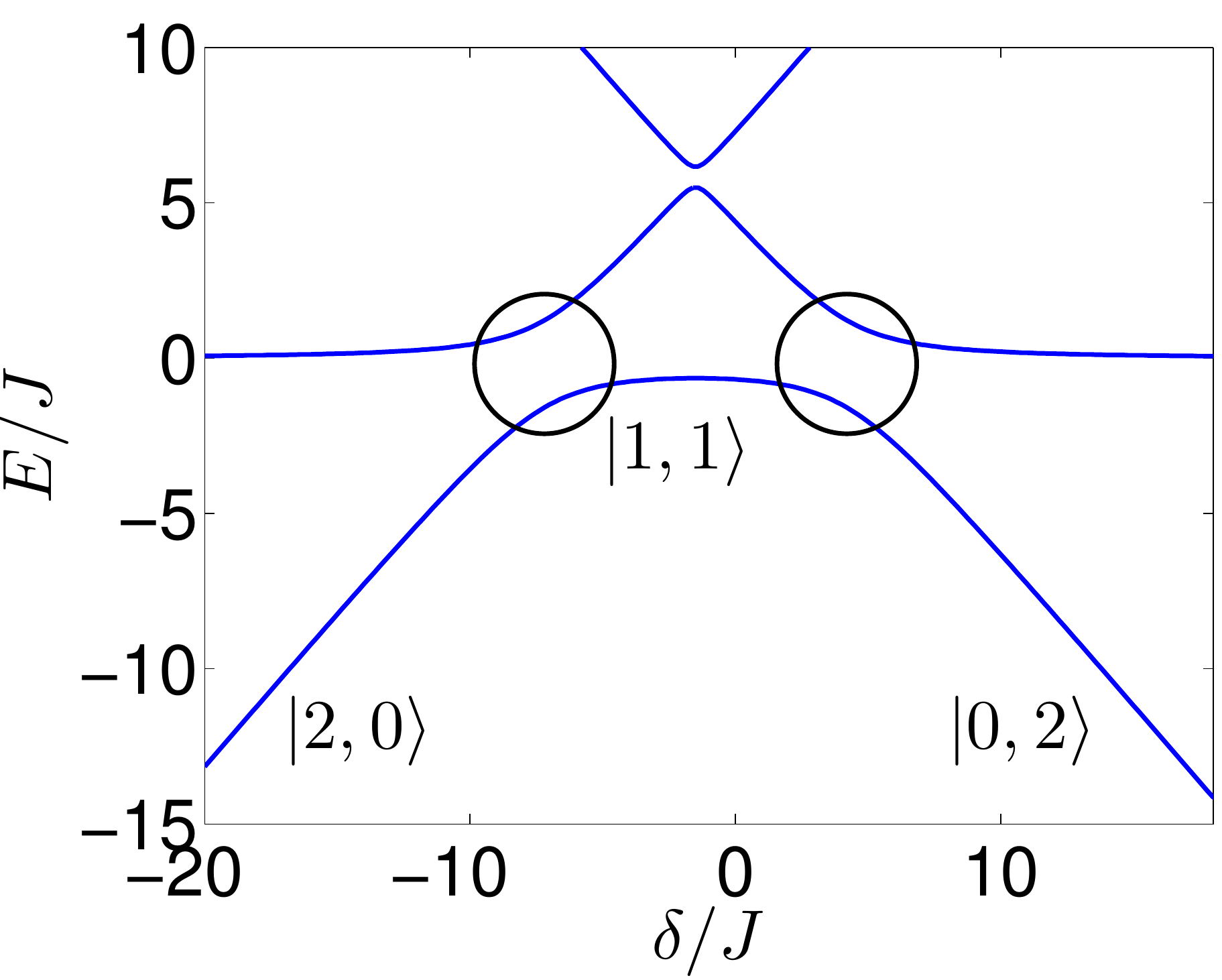}
\caption{Level structure within the Bose-Hubbard model for (left) $N=1$ and (right) $N=2$. Rapid quench to the position of the anti-crossing will induce coherent oscillations with a frequency proportional to the inverse of the energy splitting. The parameters are: $U_L/J=7$, $U_R/J=4$. }
\label{fig:BHN1N2}
\end{figure}

The phenomenon of interaction blockade in ultracold gases was first observed experimentally in an array of double wells \cite{Cheinet2008}. The simplest theoretical description is obtained in a tight-binding model, which leads to a two-mode Bose-Hubbard (or Lipkin-Meshkov-Glick model \cite{lipkin1965}) type Hamiltonian \cite{Cheinet2008} given by
\begin{align}\label{eq:Hamilt}
\hat{H}  =-&\sum_{\ell\neq \ell'}J\hat{b}_{\ell}^{\dagger}\hat{b}_{\ell'}+\sum_{\ell}\frac{U_{\ell}}{2}\hat{n}_{\ell}(\hat{n}_{\ell}-1)+\sum_{\ell}\epsilon_{\ell}\hat{n}_{\ell},
\end{align}
where $\hat{b}_{\ell}^{\dagger}$ and $\hat{b}_{\ell}$ are bosonic creation and annihilation operators with $\ell \in \{L,R\}$ denominating the left and right well, respectively, and $\hat{n}_{\ell}$ is the number operator on site $\ell$. The tunneling between the two wells is characterized by $J$, the on-site interaction coupling is $U_{\ell}$, and the on-site single-particle energy is $\epsilon_{\ell}$.

For a fixed value of $U_{\ell}/J$, avoided level crossings can be found between different eigenstates of the system when the double well is tilted, which corresponds to a relative shift of  the on-site single-particle energies $\delta \equiv \epsilon_L-\epsilon_R$. Examples for $N=1$ and $N=2$ are shown in Fig.~\ref{fig:BHN1N2}. Away from the avoided crossings, the levels simply correspond to integer occupation numbers in each well. Rapidly ramping the shift $\delta$ into the middle of any one of the avoided crossings marked by circles in Fig.~\ref{fig:BHN1N2} will initiate coherent oscillations that correspond to a single particle tunneling between the two wells.

As we will see in detail below, an interesting effect in the interaction blockade regime is that the scaling of the multiparticle energy splitting is given by the bosonic enhancement factor: 
\begin{equation}\label{eq:bef}
 \Delta E = \sqrt{N} 2J,
\end{equation}
where $\Delta E \equiv E_2-E_1$ is the energy difference between the the first-excited and ground-state energies for $N$ bosons. Note however that Eq.~\eqref{eq:bef} is only formally true in the limit 
of infinite interaction strength on one of the sites $U_{\ell}/J \to \infty$. In order to show this, we express the Hamiltonian matrix using Fock states $|n,N-n\rangle$, where $n$ is the number of bosons in the left well and $N-n$ is the remaining bosons on the right well. The full Hamiltonian matrix corresponding to Eq.~\eqref{eq:Hamilt} reads
\begin{widetext}
\begin{equation}\label{eq:hmat}
H = 
\left( 
\begin{array}{@{}c|c@{}}
   \begin{array}{@{}cccc@{}}
      \frac{U_L N(N-1)}{2}+\frac{N\delta}{2} & -J\sqrt{N} & \cdots &  0 \\
      -J\sqrt{N} & \ddots & \cdots & \vdots \\
      \vdots & \vdots & \ddots & \vdots \\
      0 & \cdots & \cdots & U_L+\frac{U_R(N-2)(N-3)}{2}+(2-\frac{N}{2})\delta
   \end{array} 
   &\begin{array}{@{}cc@{}}
      0 & 0 \\
      \vdots & \vdots\\
      \vdots & \vdots \\
      -J\sqrt{2(N-1)} & 0 
   \end{array} \\
   \hline
   \\
   \begin{array}{@{}cccc@{}}
      0 & \cdots & \cdots & -J\sqrt{2(N-1)} \\
      0 & \cdots & \cdots & 0 
   \end{array} & \begin{array}{@{}cc@{}}
   \frac{U_R(N-1)(N-2)}{2} -\frac{(N-2)\delta}{2} & -J\sqrt{N} \\ 
   -J\sqrt{N} & \frac{U_R N(N-1)}{2}-\frac{N\delta}{2}\end{array}\\
\end{array}
\right)
\end{equation}
\end{widetext}
Focussing on the lower right corner of the matrix in 
Eq.~\eqref{eq:hmat}, it becomes apparent that 
the diagonal elements become degenerate when $\delta=U_R(N-1)$. Under this condition 
the energy of a state with a single boson in the left well matches or becomes resonant with the energy of the interacting bosons all being in the right well. If, additionally $U_L /J \to \infty$, the rest of the matrix decouples and, after subtracting the degenerate energy value of the diagonal elements, the low-energy Hamiltonian reduces to  the $2 \times 2$ matrix 
\begin{equation}\label{eq:hmatN}
H_{\mathrm{res},N}=
\begin{pmatrix} 0 & -J\sqrt{N} \\ -J\sqrt{N} & 0 \end{pmatrix},
\end{equation}
which has eigenvalues reproducing the energy difference of Eq.~\eqref{eq:bef}.

The eigenstates of the resonant Hamiltonian \eqref{eq:hmatN} correspond to the ground and first-excited states of the system. They  are symmetric and asymmetric superposition of the Fock states
$|0,N\rangle$ and $|1,N-1\rangle$, respectively. This means that if an initial state $|\Psi(t=0)\rangle$ is either $|0,N\rangle$ or $|1,N-1\rangle$, the dynamical behavior is expected to exhibit coherent oscillations between these two states. For example if $|\Psi(t=0)\rangle=|0,N\rangle$, the time evolution of the many-body wave function follows from
\begin{align}\label{eq:psit}
|\Psi(t)\rangle &= e^{-iHt/\hbar}|\Psi(t=0)\rangle \\ \nonumber
&= \mathrm{cos}(\sqrt{N}Jt/\hbar)|0,N\rangle - i\mathrm{sin}(\sqrt{N}Jt/\hbar)|1,N-1\rangle.
\end{align}
The state in Eq.~\eqref{eq:psit} physically represents the tunneling of a single boson between the two wells.

It should be noted that the conditions of resonance $\delta=U_R(N-1)$, and large interactions $U_L /J \to \infty$, are difficult to satisfy in symmetric double-wells where $U_R=U_L$ because the large required tilt $\delta$ would necessarily lead to a distortion of the potential. It is therefore sensible to consider an asymmetric double-well potential, as will be done in the following section.

\begin{figure}[!htb]
\includegraphics[width=0.5\columnwidth]{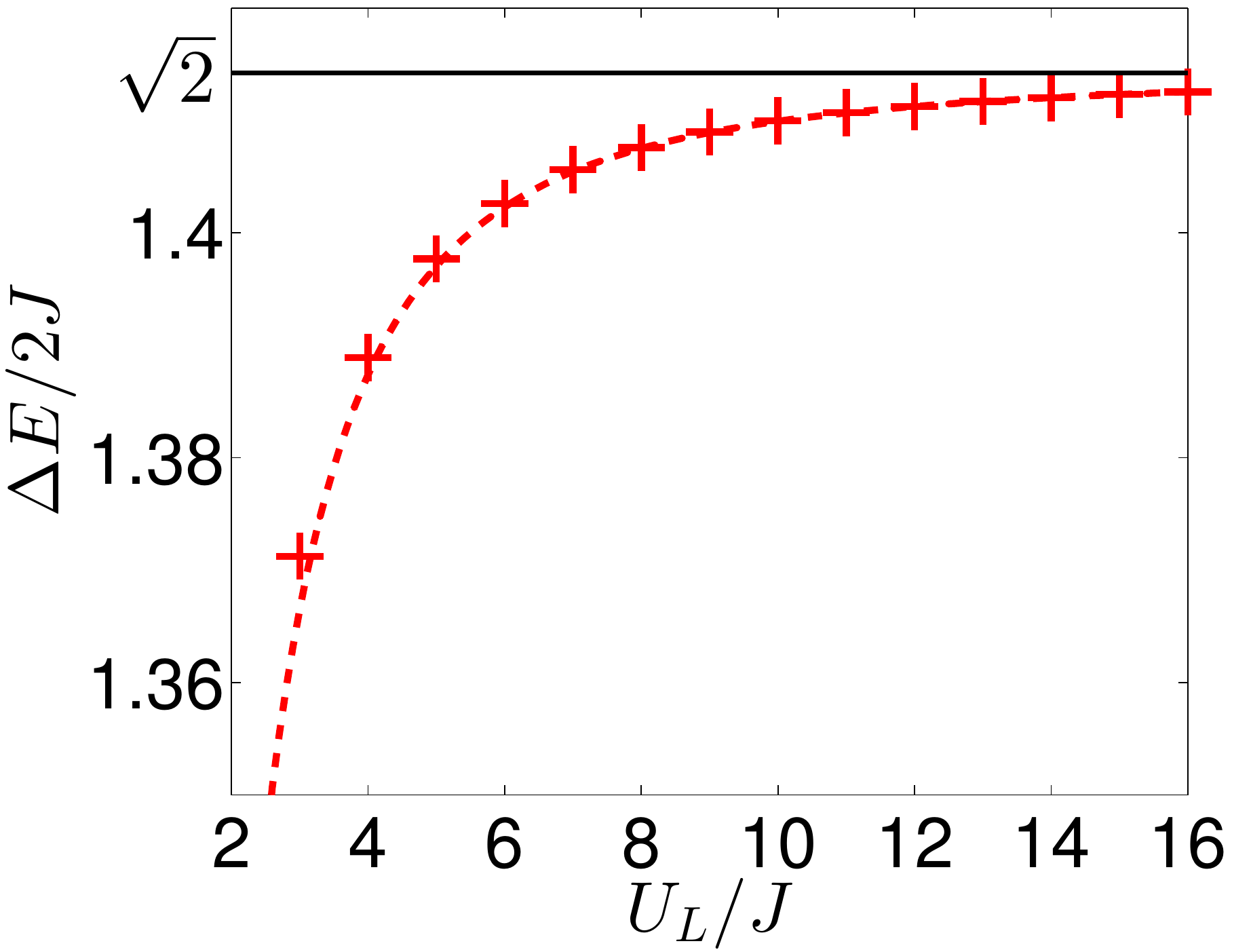}\includegraphics[width=0.5\columnwidth]{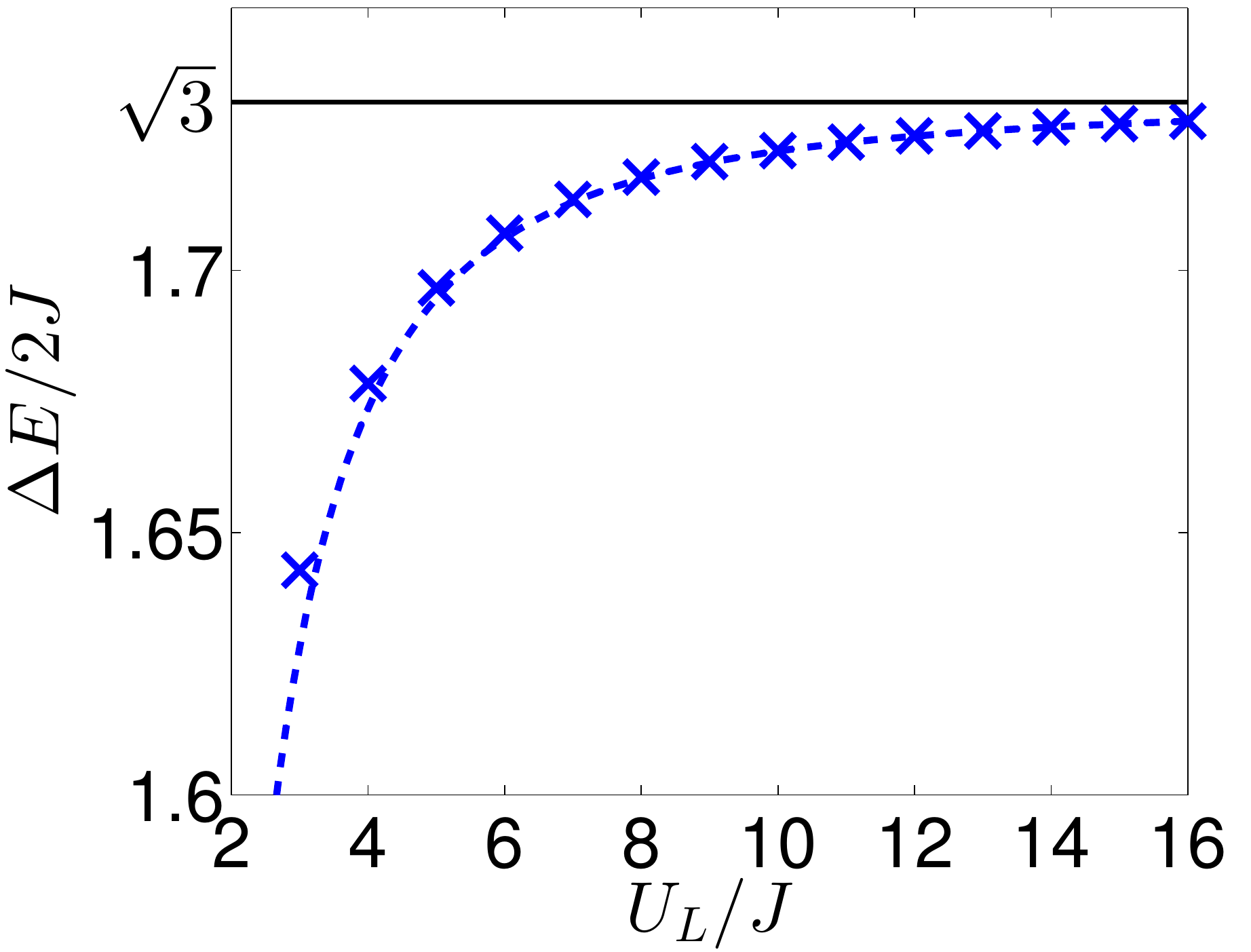}
\caption{Energy splittings at tunneling resonances involving single boson tunneling 
for $N=2$ (left) and  $N=3$ (right). Markers correspond to numerical results from exact diagonalization of the full Hamiltonian. Solid lines correspond to the expected $\sqrt{N}$ scaling in the interaction blockade limit. Broken lines denote the result of first-order perturbation in Eq.~\eqref{eq:1ord}. Other parameter: $U_R=4U_L/7$.}
\label{fig:BH} 
\end{figure}

If the conditions for the reduction to the  $2 \times 2$ matrix \eqref{eq:hmatN} are not strictly met, avoided level crossings can still be found at appropriate values of the tilt $\delta$ where coherent oscillations are possible with a frequency given by $\Delta E/2\hbar$. Experimental realisations with finite on-site interaction energy $U_{\ell}$, e.g., were reported in Refs.\ \cite{Cheinet2008,Preiss2015}.
It turns out that the blockade criteria can be ``softened'' by using finite but sufficiently strong interaction strengths $U_L/J \gg 1$, which leads to an enhancement factor very close to $\sim \sqrt{N}$ as shown in Fig.~\ref{fig:BH}.

The deviation of the energy splittings $\Delta E/2J$ from from the $\sqrt{N}$ behavior due to finite $U_L/J$ can be understood using perturbation theory as follows. The Hamiltonian matrix in Eq.~\eqref{eq:hmat} can be partitioned according to the vertical and horizontal lines and the energy eigenvalues $E$ and eigenstates can be found by solving:
\begin{equation}
H\begin{pmatrix}
  x\\
  y
 \end{pmatrix}
\equiv \left(
 \begin{array}{c|c}
  A & B \\
  \hline
  B^\dagger & H_{\mathrm{res},N}
 \end{array}
 \right)
 \begin{pmatrix}
  x\\
  y
 \end{pmatrix}
 =E\begin{pmatrix}
  x\\
  y
 \end{pmatrix},
\end{equation}
which in terms of $y$ reduces to:
\begin{equation}\label{eq:pertham}
 Ey=\left[ H_{\mathrm{res},N} + B^\dagger (E-A)^{-1} B \right]y.
\end{equation}
The second term on the right hand side of Eq.~\eqref{eq:pertham} can be treated as a perturbative correction to $ H_{\mathrm{res},N}$. Since $U_L$ is large, we can further approximate that the leading order contribution of $(E-A)^{-1}$ will come from the lowest eigenvalue of $A$, which can be approximated by $A_{mm}=U_L+U_R$. 
The energy splitting according to a first-order approximation for finite $U_{\ell}/J$ reads
\begin{equation}\label{eq:1ord}
 \Delta E/2J \approx \sqrt{N}\left[ 1 - \frac{1}{2}\left(\frac{N^2-1}{N^2}\right)\eta^2 + \mathcal{O}(\eta^4) \right],
\end{equation}
where the small parameter in the above series expansion is given by
\begin{equation}
  \eta \equiv \frac{\sqrt{N}J}{(U_L+U_R)},
\end{equation}
which is obviously small for large interaction coupling strength $U_L$ that we consider here.
Note that $\eta \to 0$ in the limit $U_L/J \to \infty$ such that we recover the expected bosonic enhancement factor $\sqrt{N}$ as per Eq.~\eqref{eq:bef}. Good agreement between the prediction of Eq.~\eqref{eq:1ord} and the numerical result from exact diagonalization is seen in Fig.~\ref{fig:BH}

The bosonic enhancement factor can be verified in experiments from measurements of the period of coherent oscillations between states representing the tunneling of a single particle between the wells \cite{Preiss2015}. For the simplest case of $N=1$, a boson will undergo coherent oscillations according to $\langle \hat{n}_L(t) \rangle = \mathrm{cos}^2(Jt/\hbar)$, such that the single-particle tunneling period is just $t^{N=1}_{\mathrm{Tun}}=h/(2J)$. 

For interacting bosons with finite but large $U_L/J$, observing coherent oscillations between $|0,N\rangle$ and $|1,N-1\rangle$ is still possible in the presence of appropriate avoided level crossings. To see this, consider the Hamiltonian $H$ with eigenvalues $\{E_k\}$ and eigenstates $\{|k\rangle\}$ wherein $|1\rangle~(|2\rangle)$ denotes the ground (first-excited) state with $E_1~(E_2)$ ground-state (first-excited state) energy. We can write the time evolution of an initial many-body state $|\Psi(0)\rangle=|n_0\rangle$ as
\begin{align}\label{eq:psitreal}
|\Psi(t)\rangle &= e^{-iHt/\hbar}|n_0\rangle \\ \nonumber
&= e^{i\Delta E t/(2\hbar)}C^{1}_{n_0}|1\rangle + e^{-i\Delta E t/(2\hbar)}C^{2}_{n_0}|2\rangle \\ \nonumber
&+\sum_{k' \notin \{1,2\}} e^{-i E_{k'} t/\hbar}e^{-i \xi t/\hbar} C^{k'}_{n_0}|k'\rangle,
\end{align}
where $C^k_{n_0}=\langle k|n_0 \rangle$ and we shift the many-body energies by $\xi = -(E_2+E_1)/2$ in order to have $E_2+\xi = \Delta E/2$ and $E_1+\xi = -\Delta E/2$.
Let us now consider an initial localized state, say for example $|n_0\rangle = |0,N\rangle$, which has a large overlap with the symmetric superposition of the two lowest eigenstates $|0,N\rangle \approx (|1\rangle + |2\rangle)/\sqrt{2}$. In addition, the corresponding antisymmetric superposition is $|1,N-1\rangle \approx (|1\rangle - |2\rangle)/\sqrt{2}$. These would mean that the overlaps are small $\{|C^{k'}_{n_0}|\} \ll 1$ between the initial state and the remaining eigenstates  $k'\notin \{1,2\}$. In this case, the rest of the terms in the last line of Eq.~\eqref{eq:psitreal} has negligible contribution and thus, the wave function will evolve as
\begin{equation}\label{eq:psitreal2}
 |\Psi(t)\rangle\approx\mathrm{cos}\left(\frac{\Delta{E}t}{2\hbar}\right)|0,N\rangle - i\mathrm{sin}\left(\frac{\Delta{E}t}{2\hbar}\right)|1,N-1\rangle.
\end{equation}
Notice how this expression resembles Eq.~\eqref{eq:psit} but with a subtle yet important distinction that in Eq.~\eqref{eq:psitreal2}, we are not approximating the Hamiltonian with the $2 \times 2$ matrix Eq.~\eqref{eq:hmatN}. But instead, we have to obtain the many-body energy splitting $\Delta E = E_2 - E_1$ from the full Hamiltonian $H$. From Eq.~\eqref{eq:psitreal2}, we find that the tunneling period can be approximated by $t^{N}_{\mathrm{Tun}}=h/(\Delta E)$. 

One of the key aspects not captured by the simple Bose-Hubbard theory is that the tunneling parameter $J$ may change when particle number is changed due to the need to adjust the tilt $\delta$ to locate the resonances. The effect becomes particularly severe in symmetric double well systems obtained in optical lattices, where $U_L \approx U_R \gg J$ as in Refs.~\cite{Cheinet2008,Preiss2015}. The inadvertent change of $J$  was considered the reason for observing a faster oscillation frequency than the expected $\sqrt{2}$ enhancement for doubly occupied sites in Ref.~\cite{Preiss2015}. 

\section{Interaction blockade in asymmetric double well}

\begin{figure}[!htb]
{\includegraphics[width=0.8\columnwidth]{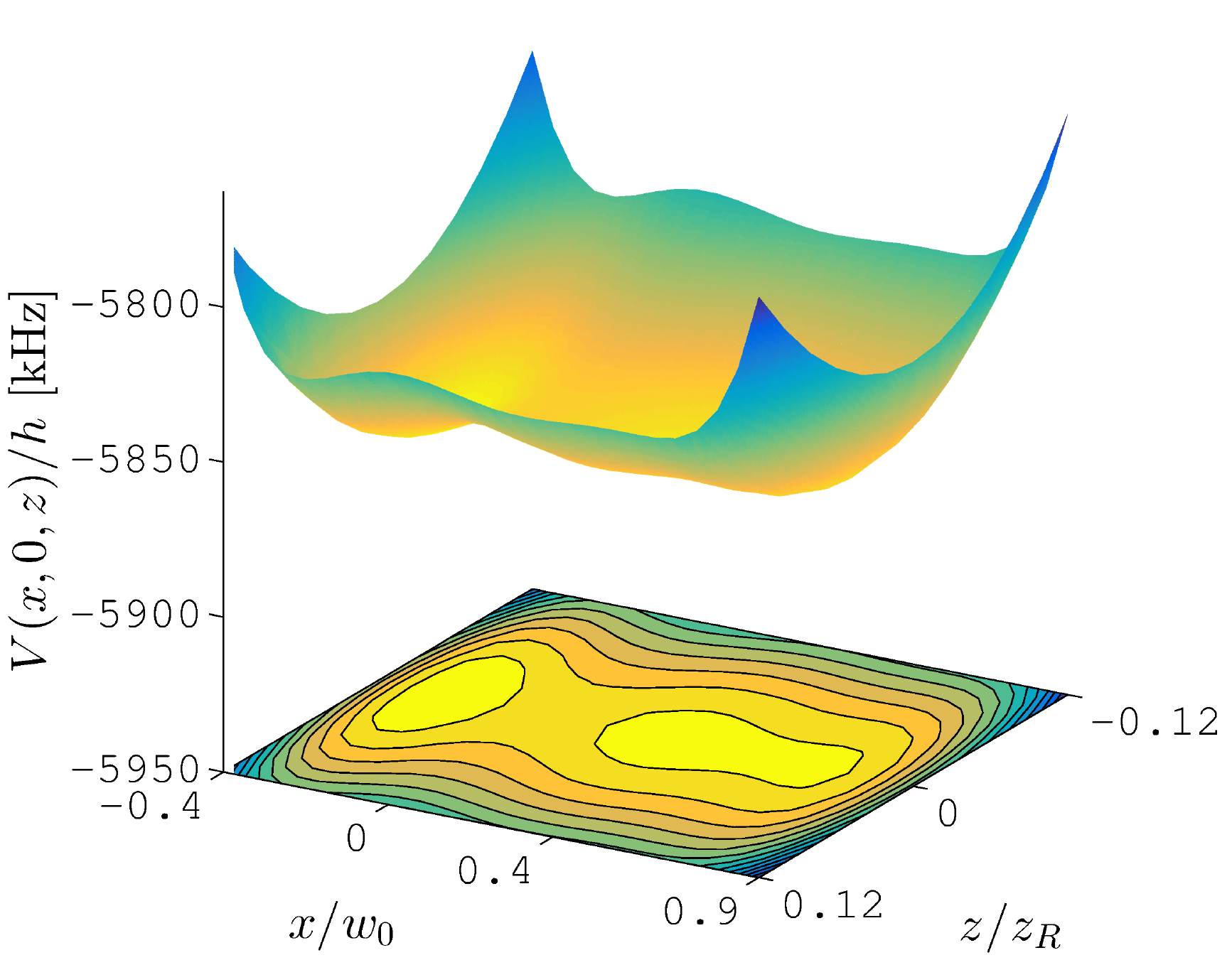}}
\caption{Asymmetric double well optical tweezer at the $x-z$ plane. Shown is the potential $V(r)$ of Eq.~\eqref{eq:pot} formed with $P=3$ laser beams for parameter values: $\{V_i/h\}[\mathrm{kHz}]=\{ 4659.775,~4137.962,~4585.886 \}$ and $\{x_i/w_0\}=\{-0.658,~0.264,~1.176\}$.
}
\label{fig:3Dpot} 
\end{figure}

We now show that one can use an asymmetric double well to observe the $\sqrt{N}$ bosonic enhancement factor according to Eq.~\eqref{eq:bef}.
Specifically, we propose to disentangle the single-particle coupling parameter $J$ from the tilt $\delta$
by using an asymmetric double well potential formed from the optical dipole potential of three laser beams (optical tweezers) far red-detuned off  the closest atomic resonance such that they provide conservative attractive potentials for the atoms proportional to the local light intensity. Two closely spaced laser beams will combine to form a wide (right) well while the other laser beam will serve as a narrow (left) well separated by a small barrier. An example of such configuration is depicted in Fig.~\ref{fig:3Dpot}. We further suggest to probe the avoided level crossings by changing the intensity of the laser beam farthest (right-most) from the effective barrier. Tuning this intensity will almost only affect the right-most region of the potential, effectively adjusting the bias $\epsilon_R$, while leaving the barrier region responsible for the single-particle coupling $J$ mostly unaffected. 
Another advantage of using an asymmetric configuration is that the effective on-site interaction energy in the wide well will be naturally smaller than the other well, i.e., $U_R < U_L$. Thus, the resonance condition can be achieved with relatively little change on the depth of the right-most potential when compared to the symmetric double well case.

\subsection{Optical tweezer potential}\label{sec:tweezer}

Here, we discuss one possible experimental realization of such an asymmetric double well trap. In particular, we consider a superposition of multiple 3D optical dipole potentials modeled by \cite{Wall2015}
\begin{align}\label{eq:pot}
 V(\mathbf{r}) = -\sum_{i=1}^P \frac{V_i}{1+\frac{z^2}{z_R^2}} \mathrm{exp} &\biggl[-\frac{2((x-x_i)^2+y^2)}{w_0^2\biggl(1+\frac{z^2}{z_R^2}\biggr)}\biggr]\\ \nonumber
 &+mgy,
\end{align}
where $m$ is the mass of the bosonic atom, $g$ is the gravitational acceleration $g=9.81~\mathrm{m/s^2}$, $P$ is the number of laser beams utilized in the trap, $\{V_i\}~(\{x_i\})$  are the depths (positions) of each beam, $w_0$ is the beam waist, and $z_R=\pi w_0^2/\lambda$ is the Rayleigh range, with $\lambda$ being the wavelength of the laser. The double-well shape in the 3D potential Eq.~\eqref{eq:pot} is introduced along the $x$-direction. We also include the gravitational potential along the $y$-direction for the sake of completeness even though the overall effect is relatively small. 
In the following discussion, we have used experimentally relevant parameters for the mass $m$ of a $^{85}\mathrm{Rb}$ atom, the beam waist $w_0=1.015~\mu\mathrm{m}$, and the laser wavelength $\lambda = 1.064~\mu\mathrm{m}$. Since the background $s$-wave scattering length for $^{85}\mathrm{Rb}$ is negative \cite{Altin2010,Chin2010}, we assume that the scattering length can be tuned via Feshbach resonance and instead use $a_s=5.45$ nm.
Although not presented here, we have checked that the main findings of our work are robust against changing the isotope by repeating our calculations for $^{87}\mathrm{Rb}$.

\begin{figure}[!htb]
\includegraphics[width=0.9\columnwidth]{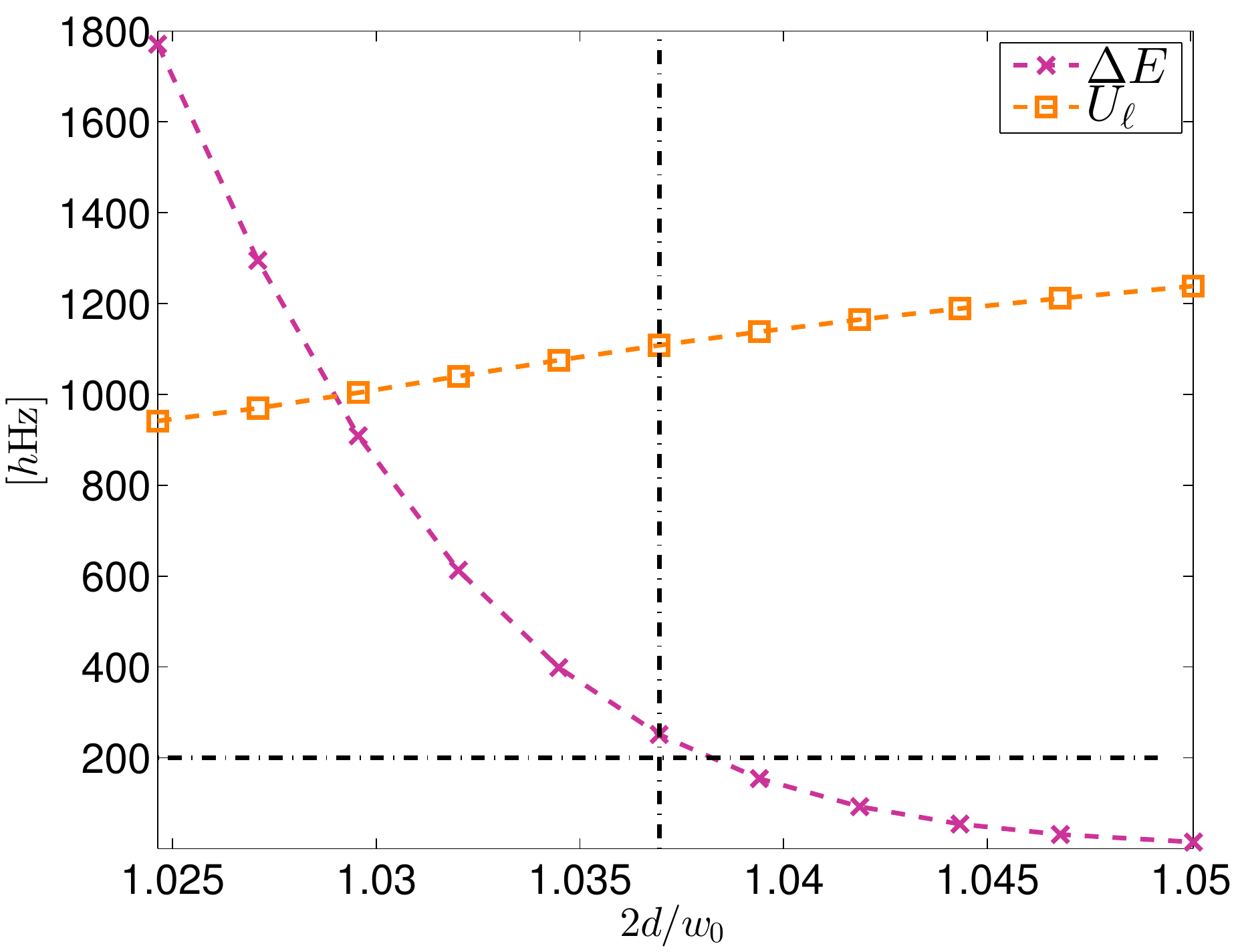}
\caption{Single-particle energy splitting $\Delta E$ for $N=1$ and intra-well interaction energy  $U_{\ell}=U_L=U_R$ of a symmetric double with two optical tweezer beams as a function of the beam separation $d$ with $V_0/h=5~\mathrm{MHz}$. The horizontal dashed-dotted line represents the minimum energy splitting for experimentally observable single-particle tunneling.}
\label{fig:sep} 
\end{figure}

The parameter space becomes fairly complex when we consider at  three or more optical tweezer beams in the system because of the different possible combinations of depths $\{V_i\}$ and positions $\{x_i\}$ of the beams. To narrow down the search for suitable trap parameters we first consider a symmetric double well with a potential along the $x$-axis ($y=z=0$) given by
\begin{align}
 V(x,0,0) = -V_0 \biggl[ &\mathrm{exp}(\frac{-2(x+d)^2}{w_0^2}) \\ \nonumber
  &+ \mathrm{exp}(\frac{-2(x-d)^2}{w_0^2})\biggr] .
\end{align}
Note that at a separation of $d=w_0/2$ the two wells will merge. In practice, there are two important considerations that need to be taken care of: (i) reaching the interaction blockade regime of $U \gg J$, and (ii) the energy splitting is large enough to be experimentally observed. For the first condition, we choose $U/J \sim 10$. For the second condition we choose a minimum energy splitting of $\Delta E/h \sim 200~\mathrm{Hz}$, which corresponds to $5~\mathrm{ms}$ of oscillation period, being motivated by experimental requirements.
The effective Bose-Hubbard parameters $(J,U)$ are calculated following the prescriptions of Ref.~\cite{Wall2015} with further details given in App.~\ref{appen:b}.
We find that a suitable choice of $V_0/h=V_1/h=5~\mathrm{MHz}$ satisfies both conditions. In Fig.~\ref{fig:sep}, we show the single-particle energy splitting $\Delta E$ and interaction energies $U_{\ell}$ as a function of the double well separation.

The same two constraints for the symmetric configuration will also apply for the asymmetric double well when exploring the parameter space. We impose an additional criterion for the separation between the second and third laser beams: it must be separated as far as possible but the single-particle functions must still be delocalized between the combined well. In practice, the aforementioned conditions will effectively fix the spacings $\{x_i\}$ and the intensities of the beams forming the wide well $\{V_i\}$. 

\subsection{Few particle quantum simulations}\label{sec:blockade}

\begin{figure}[!htb]
\includegraphics[width=0.5\columnwidth]{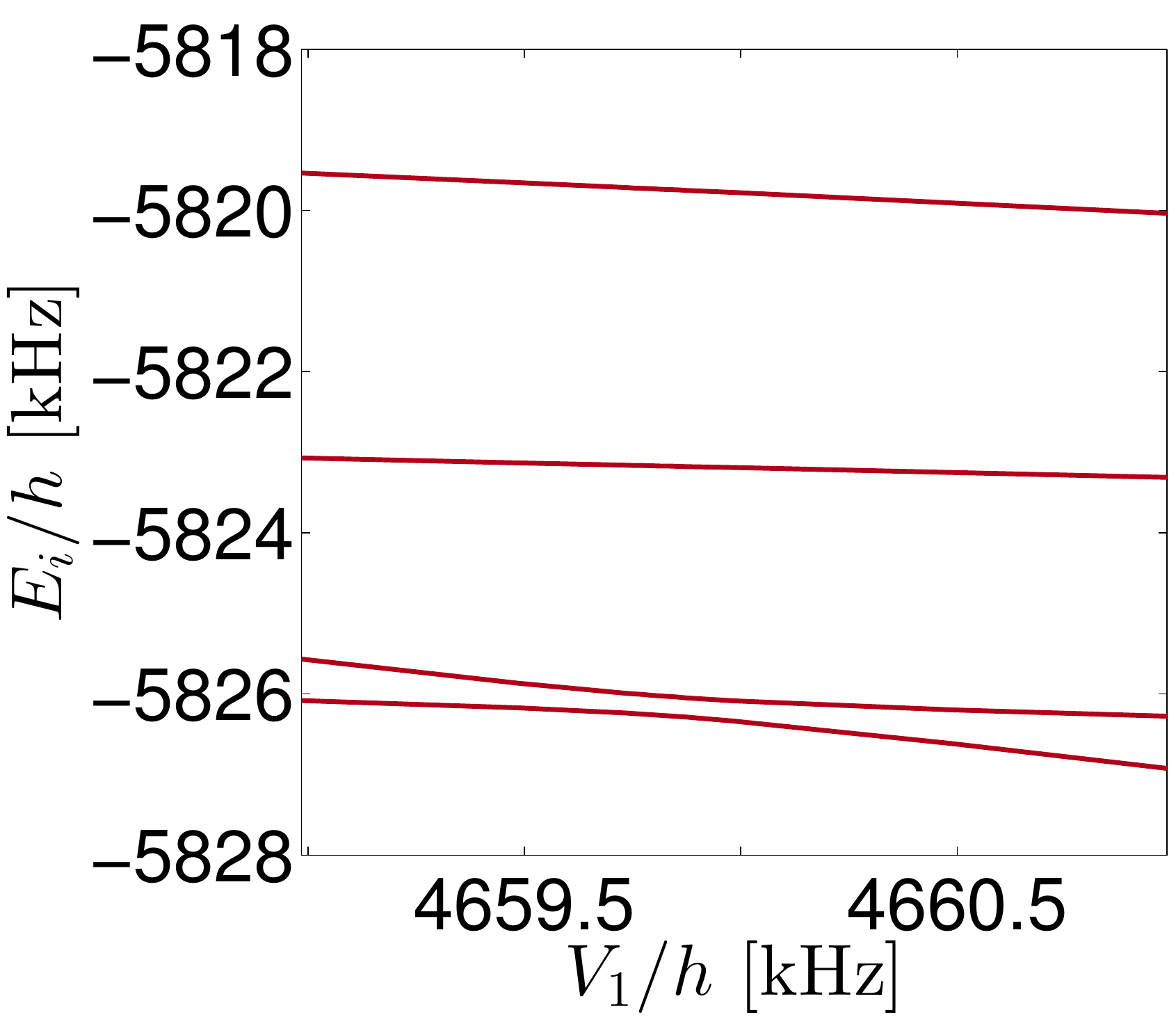}\includegraphics[width=0.5\columnwidth]{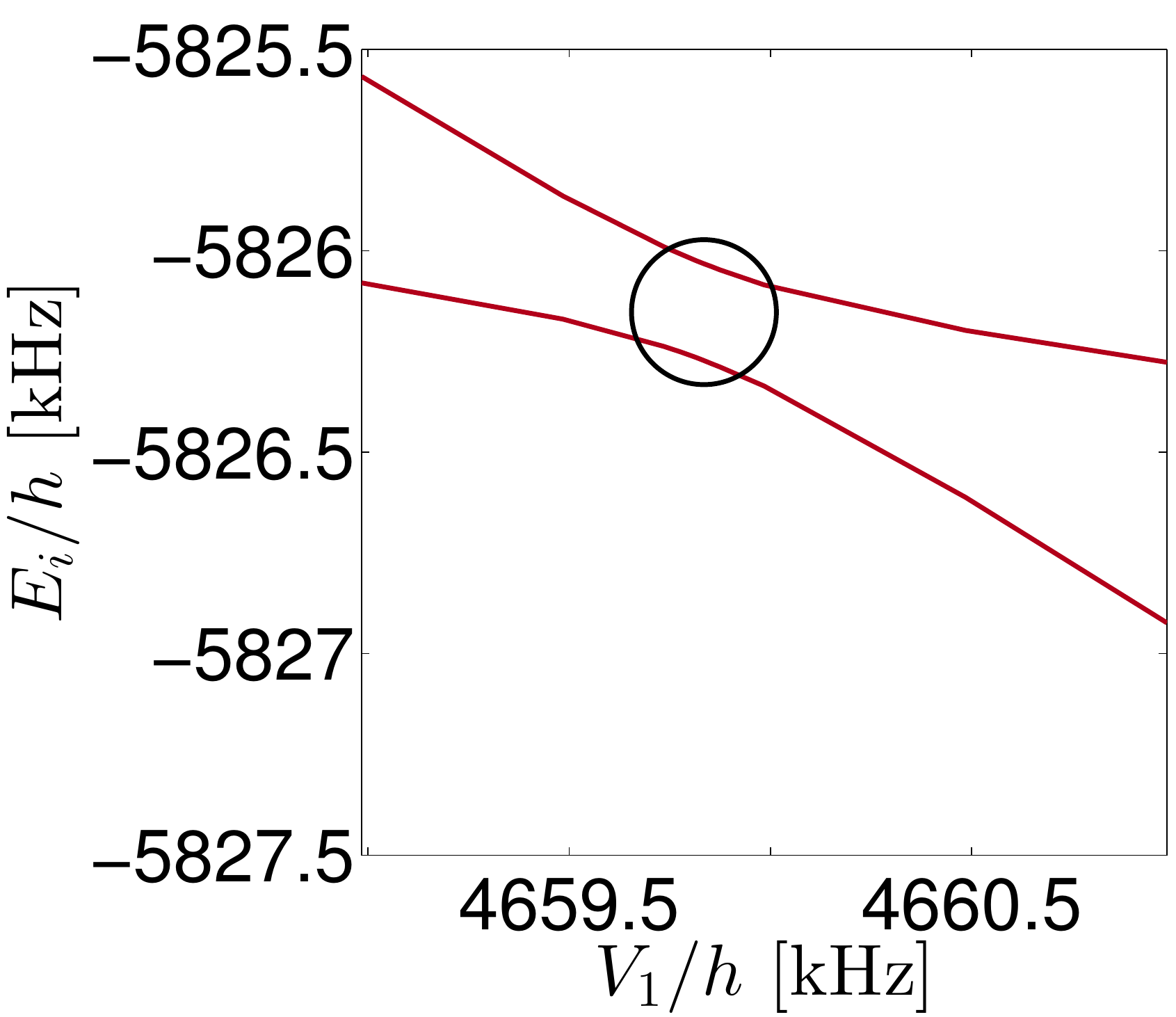}
\caption{Single-particle energies in the asymmetric double well trap  as a function of the left-most laser beam depth $V_1$: (left) First four single-particle energies and (right) Zoom-in of to the ground and first excited energies. The single-particle resonance identified from an avoided level crossing is indicated by the circle.}
\label{fig:spen} 
\end{figure}

To illustrate our suggested scheme, we choose a particular set of parameters $\{x_i,V_i\}$. Afterward, the tunneling resonance for a single boson is located by tuning the depth of the narrow well $V_1$. The single-particle energies as a function of $V_1$ are shown in Fig.~\ref{fig:spen}. For the asymmetric double well, the single-particle energy splitting between the two lowest states is $\Delta E/h = 2J/h = 240.14~\mathrm{Hz}$, which leads to a tunneling period of $t_{\mathrm{Tun}}=4.16~\mathrm{ms}$. It can also be seen in the left panel of Fig.~\ref{fig:spen} that the second- and third-excited states are well separated from the ground and first-excited states. 
This motivates us to use the MCTDHB scheme \cite{AlonEtAl2008} for fully 3D numerical simulations when $N>1$. The MCTDHB method expands a multi-particle wave function in an occupation number basis constructed from a small number $M$ of optimised orthonormal single-particle functions.
A short description of the MCTDHB method and additional details on the numerical convergence can be found in Appendix~\ref{appen:a}. An alternative method is to diagonalize the corrresponding Bose-Hubbard Hamiltonian for $N$ bosons using the effective Bose-Hubbard parameters calculated according to the prescription of Ref.~\cite{Wall2015} as the beam depth is being varied. The Bose-Hubbard parametrization for the asymmetric double well is discussed briefly in Appendix~\ref{appen:b}. In this context, the full $(N+1) \times (N+1)$ Hamiltonian matrix is constructed and diagonalized in the Fock state basis.

For the MCTDHB calculations, the three-dimensional nature of the system requires a careful treatment of the short-range interaction 
\begin{equation}
 \hat{U}(\mathbf{r}_i-\mathbf{r}_j) = \Lambda\delta(\mathbf{r}_i-\mathbf{r}_j),
\end{equation}
where $\Lambda$ is the renormalized coupling constant, used in our simulations (see Refs.~\cite{Rontani2008,Bolsinger2017} and references therein). As discussed in Appendix~\ref{appen:a}, the main convergence parameter in MCTDHB is the number of single-particle modes $M$ used to represent the MCTDHB many-body wave function. To this end, for each choice of $M$, we follow the renormalization procedure utilized in Refs.~\cite{Rontani2008,Armstrong2012,DAmico2015} when renormalizing the interaction strength with respect to the exact ground-state energy of interacting bosons in a 3D harmonic oscillator \cite{BuschEtAl1998} solved using the same spatial grid as the asymmetric double well potential. Another way of modelling short-range interactions is by using finite-range model potentials such as a Gaussian interaction potential as in Ref.~\cite{Bolsinger2017}. But as we shall see below for $N=2$ bosons, renormalizing the strength of the contact potential is already sufficient to get good agreement between the results of the MCTDHB simulations and the Bose-Hubbard model with the unrenormalized contact parameter $4\pi\hbar^2 a_s/m$.

\begin{figure}[!htb]
\includegraphics[width=0.5\columnwidth]{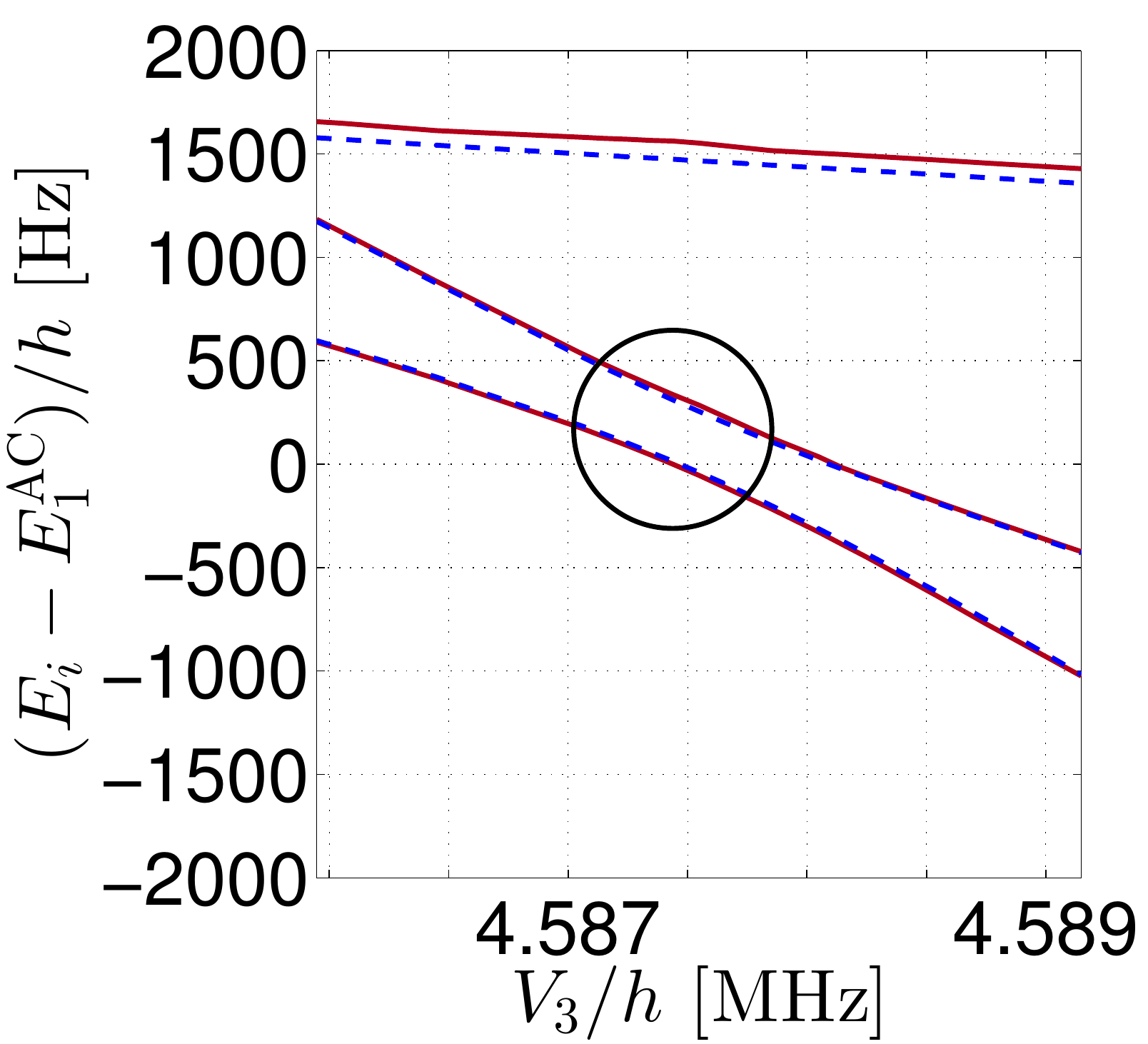}\includegraphics[width=0.48\columnwidth]{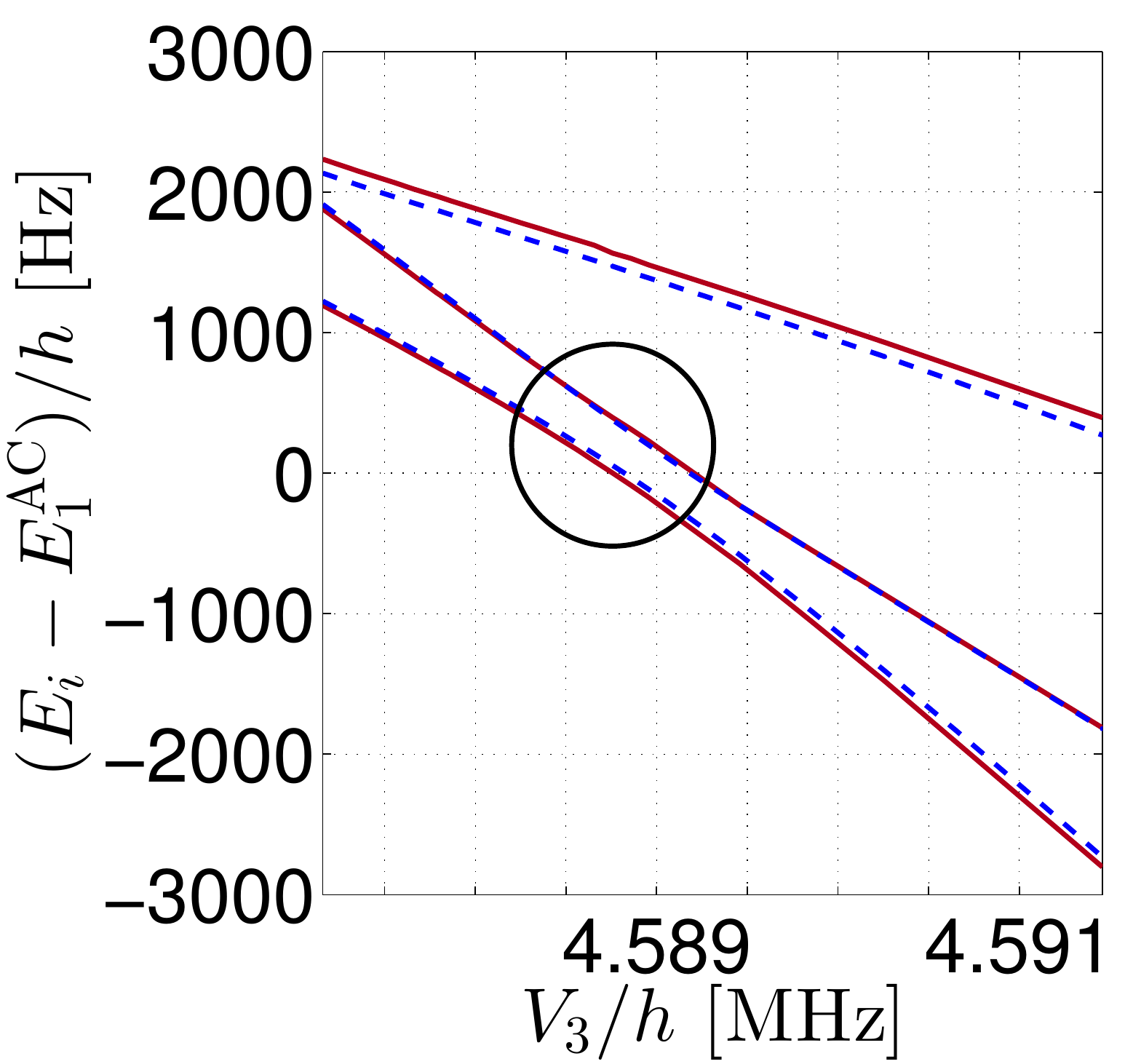}
\caption{Ground, first-, and second-excited states as a function of $V_3$ for the asymmetric double well for (left) $N=2$ and (right) $N=3$.  The reference ground-state energies are $E_1^{\mathrm{AC}}/h=-11.653~\mathrm{MHz}$ for $N=2$ and $E_1^{\mathrm{AC}}/h=-17.480~\mathrm{MHz}$ for $N=3$. Solid lines correspond to MCTDHB results. Broken lines correspond to diagonalization of the Bose-Hubbard Hamiltonian. Resonances are indicated in circles.}
\label{fig:EN2} 
\end{figure}

Now that the value of $V_1$ is fixed by the resonance condition for $N=1$, we implement our protocol of varying $V_3$ to identify the avoided level crossing for $N>1$. The results for the multi-particle eigenenergies are shown in Fig.~\ref{fig:EN2}. We have obtained energy splittings at the avoided level crossing of $\Delta E/h=336.46~\mathrm{Hz}$ for $N=2$ and $\Delta E/h=395.62~\mathrm{Hz}$ for $N=3$.
Interestingly, the two lowest multi-particle energies from the effective Bose-Hubbard model are consistent with the corresponding MCTDHB results for the parameter space explored here. The deviation seen for the second-excited state energy can be explained by the fact that in our MCTDHB calculations only the ground or the first-excited state is variationally optimized and thus, the second-excited state energy is expected to be higher than the exact energy.
We can then infer that the renormalization scheme from Ref.~\cite{Rontani2008} and the MCTDHB method are both applicable for characterizing the energy gap between the two lowest energy states for the system considered here.

\begin{figure}[!htb]
\includegraphics[width=0.5\columnwidth]{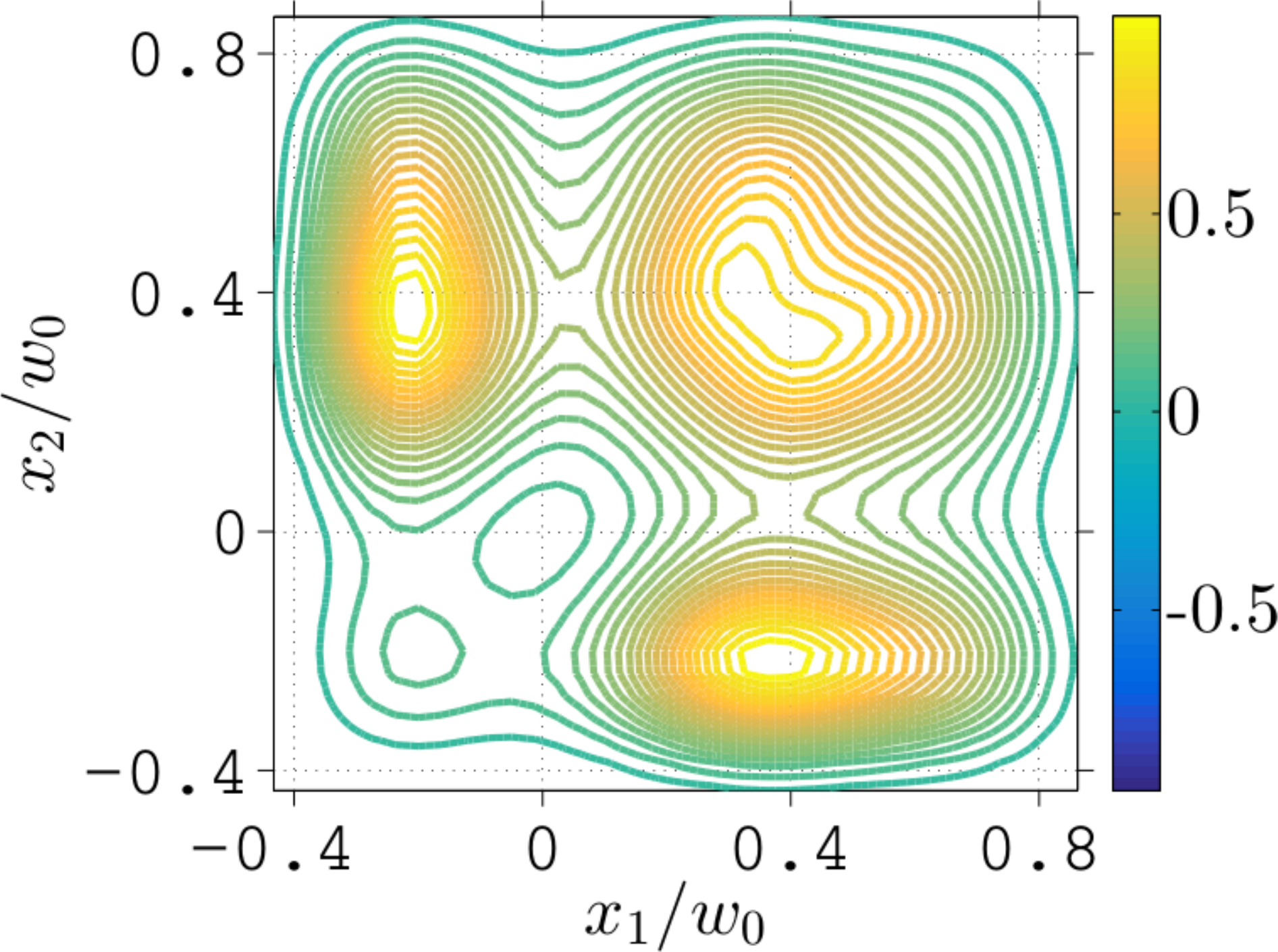}\includegraphics[width=0.5\columnwidth]{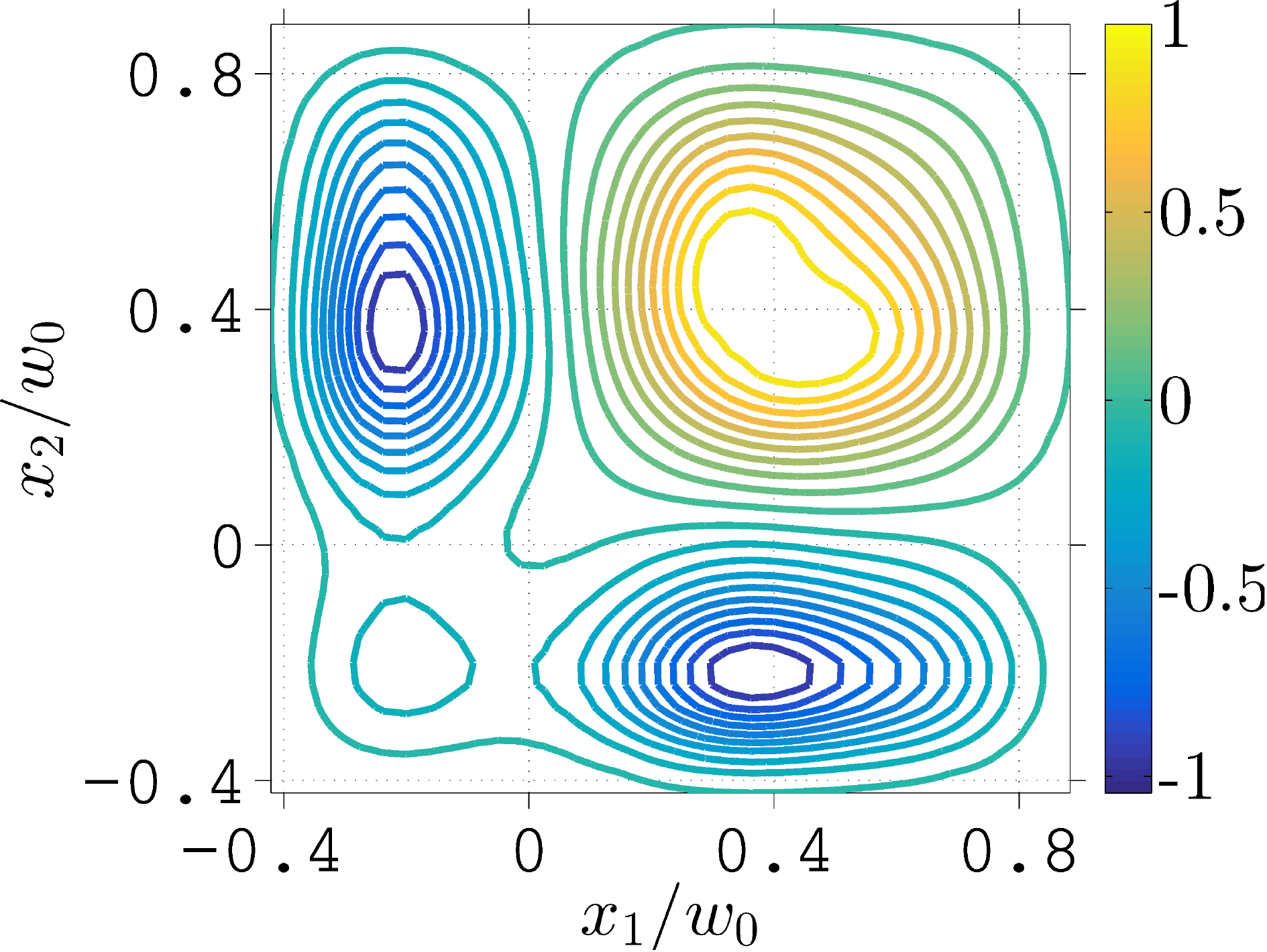}
\caption{Slice of the two-particle wave function for $N=2$ at $y_1=y_2=y_3=z_1=z_2=z_3=0$ as a function of $x_1$ and $x_2$. Left: ground state; Right: first excited state.}
\label{fig:wavN2} 
\end{figure}

\begin{figure}[!htb]
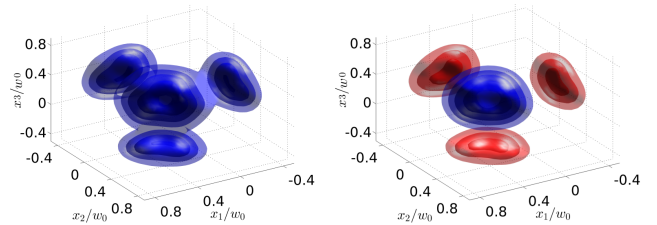

\includegraphics[width=0.5\columnwidth]{N3_Rb85_GS_wavefunc}\includegraphics[width=0.5\columnwidth]{N3_Rb85_ES_wavefunc}
\caption{Three-particle wave function for $N=3$ at $y_1=y_2=y_3=z_1=z_2=z_3=0$. The blue surfaces correspond to positive values while the red surfaces correspond to negative values of the wave function. (left) ground state and (right) first-excited state.}
\label{fig:wavN3} 
\end{figure}

Due to the three-dimensional nature of the potential, it is imperative to ensure that the avoided level crossings in Fig.~\ref{fig:EN2} correspond to the relevant states and excitations along the axis of asymmetry.
To this end, we show plots of the two-body wave function as depicted in Fig.~\ref{fig:wavN2}. In these plots, it can be seen that the first-excited state has a node along the $x$-axis at the position where the effective barrier has a maximum. 
Features of the wave function of the first-excited state shown in the right panel of Fig.~\ref{fig:wavN2} are insightful in understanding the nature of the excited states at the avoided level crossing. On one hand, the positive amplitude of the wave function for positive values of $x_1$ and $x_2$ (upper-right quadrant) describes a state of two bosons sitting on the wide well. On the other hand, the negative amplitude in the remaining domain of negative $x_1$ with positive $x_2$ and vice versa depicts a state with one boson in each well. Therefore, this excited state $|\Psi_1\rangle$ represents the antisymmetric superposition of the localized wave functions $|\Psi_1\rangle = (|0,2\rangle - |1,1 \rangle)/\sqrt{2}$.
A similar analysis can be made for the ground-state wave function $|\Psi_0\rangle$ shown in the left panel of Fig.~\ref{fig:wavN2}, leading to the conclusion that the ground state of the system corresponds to the symmetric superposition state $|\Psi_0\rangle = (|0,2\rangle + |1,1 \rangle)/\sqrt{2}$.
Thus, the states which physically represent a single boson tunneling from one well to the other are manifested in the two lowest eigenstates of the system.
In a similar fashion, we can also look at the corresponding three-body wave function for $N=3$. The important features of the isosurface plots shown in Fig.~\ref{fig:wavN3} can be regarded as simple three-dimensional (corresponding to three bosons $\{x_1,x_2,x_3\}$) extension of the contour plots in Fig.~\ref{fig:wavN2}. We see that the two lowest energy states obtained are the relevant anti-crossing states in the numerical simulation for $N=3$.

\begin{figure}[!htb]
\includegraphics[angle=-90,width=0.90\columnwidth]{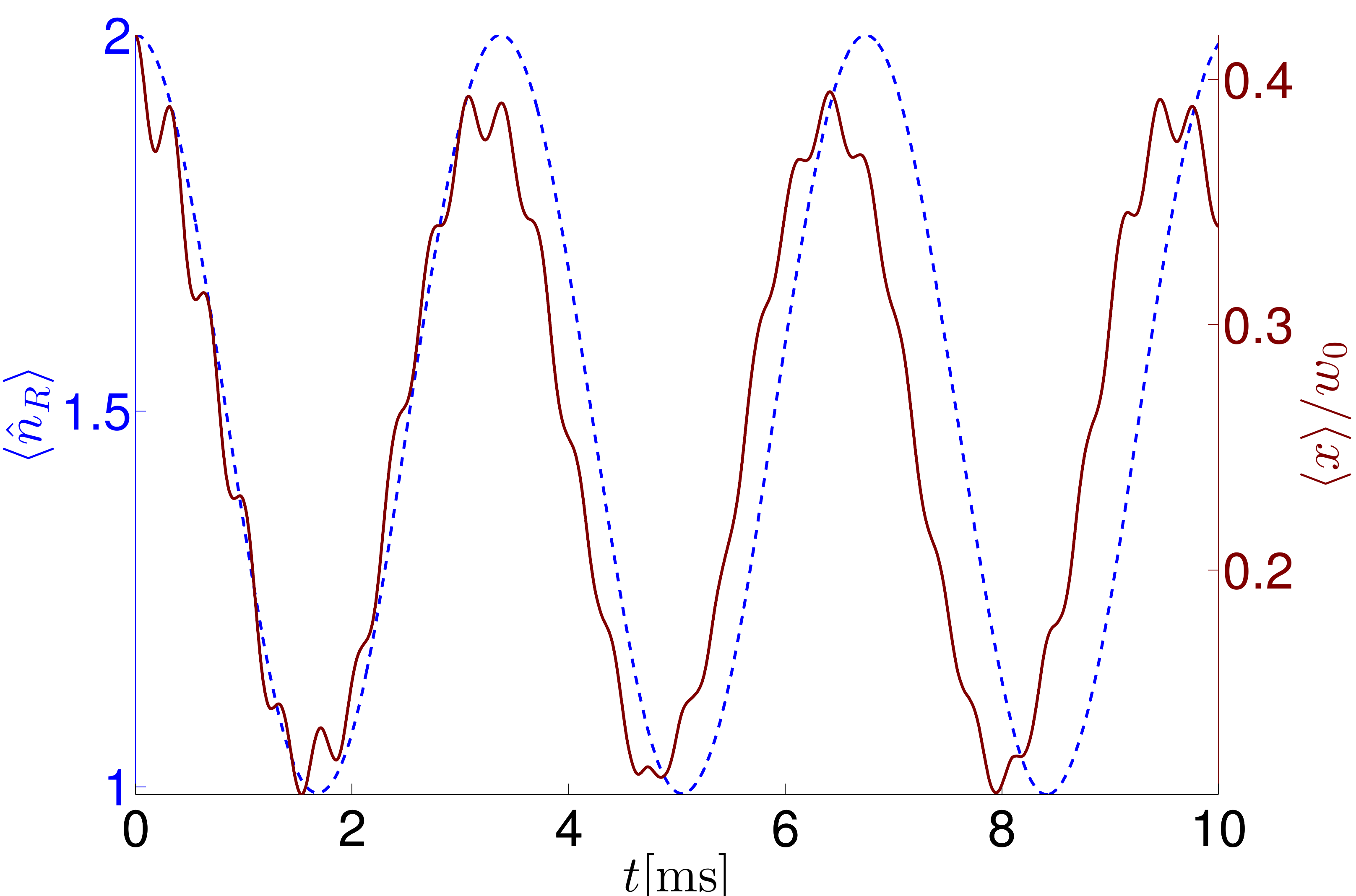}
\caption{Dynamics for $N=2$ following a quench to the position of the avoided crossing from the ground state at the largest value of $V_3$ in Fig.~\ref{fig:EN2}. Dashed-curve (left y-axis) denotes the dynamics of the expectation value of mode occupation number in the right (wide) well $\langle \hat{n}_R \rangle$ within the Bose-Hubbard model. Solid curve (right y-axis) depicts the dynamics of the average center-of-mass position in the axis of asymmetry $\langle x \rangle / w_0$ calculated from MCTDHB.}
\label{fig:N2evo} 
\end{figure}

In order to ensure that the dynamics of an initial localized state will be restricted to the two-state subspace at the avoided level crossing, we must choose an initial state with strong overlap with the symmmetric superposition of two lowest few-body eigenstates $|1\rangle$ and $|2\rangle$ at the resonance. This can be quantified by the fidelity $\mathcal{F}=|\langle \Psi_R |\Psi_0 \rangle|^2$ where $|\Psi_R\rangle = (|1\rangle + |2\rangle)/\sqrt{2}$. As an example, let us use the ground-state wave functions at the largest value of $V_3$ in Fig.~\ref{fig:EN2} as the initial state $|\Psi_0\rangle$. For $N=2$, the fidelity of this initial state with the symmetric superpositions at the avoided crossing $|\Psi_R\rangle$ is $\mathcal{F}=0.9439$. While for $N=3$, the corresponding fidelity is $\mathcal{F}=0.9306$. These numbers suggest that this procedure is viable and the subsequent coherent oscillations will be dominated by the two lowest few-body energy states. Further details on how to calculate fidelities between two MCTDHB wave functions are discussed in Appendix~\ref{appen:c}.
We now proceed to a numerical demonstration of the time evolution of relevant observables for $N=2$ following a trap quench from the ground state of the largest $V_3$ considered in Fig.~\ref{fig:EN2} to the position of the avoided level crossing. The ensuing dynamics is expected to exhibit coherent oscillations with a period associated to the numerically obtained few-body energy splitting $\Delta E$ because of the large fidelity ($\mathcal{F}=0.9439$) between the initial state and the symmetric superposition of the two lowest energy states at the resonance. Indeed, we observe in Fig.~\ref{fig:N2evo} that both Bose-Hubbard and MCTDHB results exhibit oscillatory behavior of relevant observables, which are the mode occupation number in the wide well $\langle \hat{n}_R \rangle$ for the Bose-Hubbard model and the position of the center-of-mass $\langle x \rangle / w_0$ for the MCTDHB calculation. The presence of small amplitude oscillations in the MCTDHB results can be explained by the fact that the fidelity is not perfect. Nevertheless, the oscillation frequency of the center-of-mass position is consistent with the numerically calculated $\Delta E$ for $N=2$. The difference in oscillation frequency when compared to the Bose-Hubbard results, which is about $\sim 10\%$, can be accounted for by the slight mismatch between the energy splittings obtained from the two methods. But more importantly, the time-dependent simulations presented in Fig.~\ref{fig:N2evo} corroborate the validity of the time-independent arguments based on the avoided level crossings.

\begin{figure}[!htb]
\includegraphics[width=0.9\columnwidth]{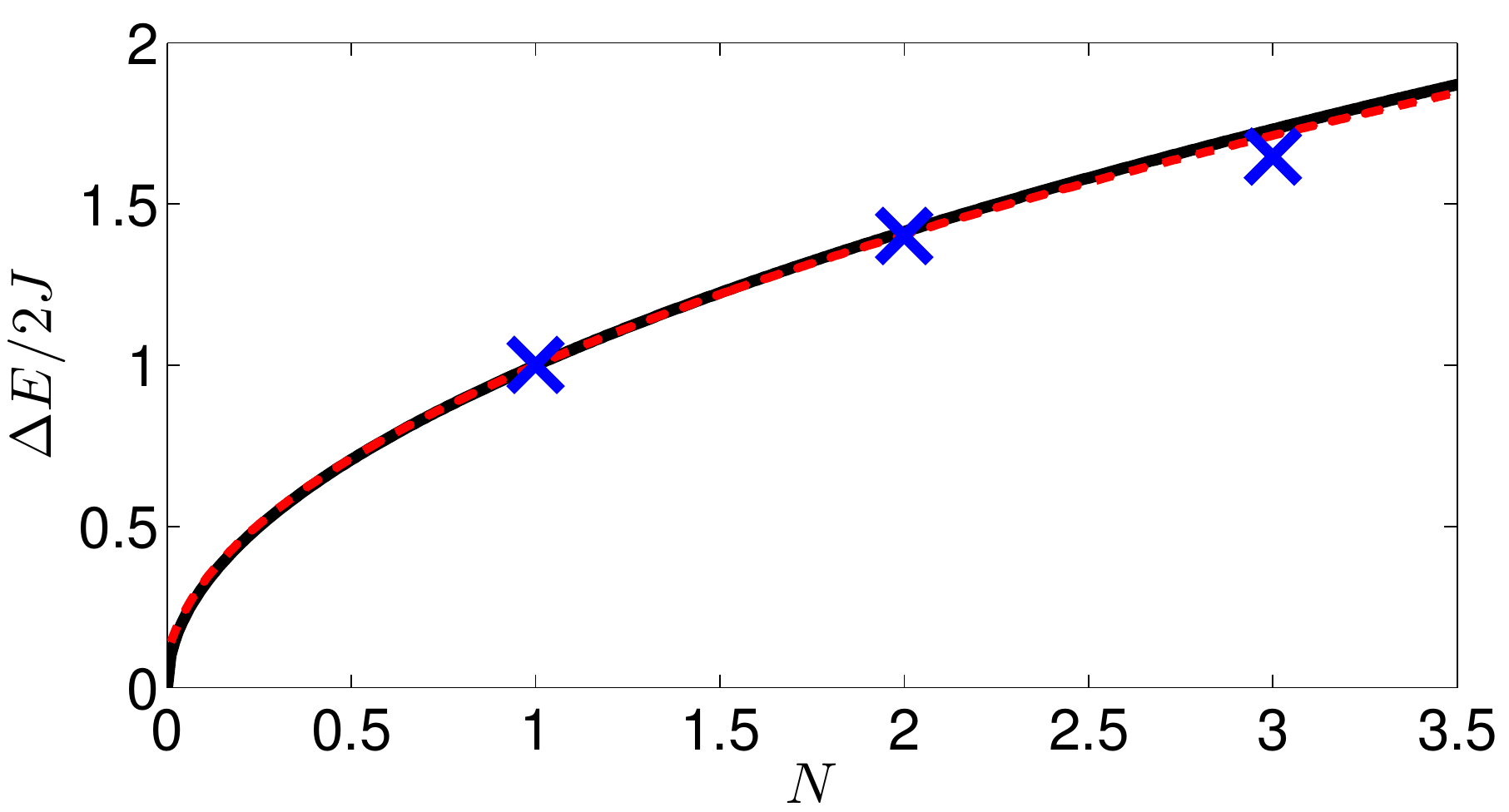}
\caption{Multiple particle energy splittings as a function of $N$. Solid line corresponds to the expected $\sqrt{N}$ dependence. The dashed line corresponds to the first-order approximated splitting given by Eq.~\eqref{eq:1ord} for an effective Bose-Hubbard parameter of $U_L/J \approx 7$ and $U_R/J \approx 4$. Blue crosses correspond to the numerical results at the positions of resonance indicated by the circles in Figs.~\ref{fig:spen} and \ref{fig:EN2}.}
\label{fig:delEN2N3} 
\end{figure}

Finally, we compare the energy splittings for $N=2$ and $N=3$ with that for $N=1$. Indeed, it can be seen from Fig.~\ref{fig:delEN2N3} that our numerical results for $\Delta E/2J$ closely follow the expected $\sqrt{N}$-behavior. For $N=2$, we can directly compare our result $\Delta E/2J=1.40$ with previous experimental observation \cite{Preiss2015}, which is $\Delta E/2J=2.37$. This clearly illustrates that our proposal of using an asymmetric double well should be able to capture the expected tunneling period in the interaction blockade regime, at least for the number of bosons presented here.

\section{Discussion and Outlook}\label{sec:disc}

In this work, we have proposed a way to experimentally observe the $\sqrt{N}$ bosonic enhancement factor for the energy splitting in the interaction blockade regime of few bosons. Our suggestion is to use an asymmetric double well configuration made from superimposing single-well potentials. This setup allows for tuning of energy levels by varying the depth of the well farthest away from the effective barrier. In doing so, the effect on the barrier that separates the left and the right wells becomes minimal as the avoided level crossing is being probed.
To test our claim, we have done a series of numerical experiments using 3D optical tweezers and obtain the relevant energy splittings. We have shown that the $\sqrt{N}$ behavior of the energy gap between the tunneling resonance states manifests for $N=2$ and $N=3$. While good agreement can be seen for $N=2$, there is already a slight deviation for the $N=3$ results from the expected $\sqrt{N}$ prediction. Unfortunately, the correction according to Eq.~\eqref{eq:1ord} for finite $U_L/J$ does not quantitatively capture the apparent decrease in the energy level splitting for $N=3$ as seen in Fig.~\ref{fig:delEN2N3}. The nature of this discrepancy can be studied in future works. In general, one might expect larger deviations for increasing $N$ because the depth of the well farthest from the barrier needs to be tuned more in order to locate the resonance. Thus, probing the avoided level crossing for larger $N$ might have more pronounced effects on the deformation of the effective barrier. 
The ideas in our proposed scheme can be extended to a system comprising of more laser beams than the three beams considered in this work. Of course, the complexity of the parameter space increases but in exchange, this consideration can improve the possibility of studying coherent transport phenomena without altering the tunneling barrier in systems with larger $N$. 

In terms of experimental implementation, the multi-well potential of Eq.~\eqref{eq:pot} can be realized by using for example spatial light modulator \cite{Barredo2016} or acousto-optical modulator \cite{Kaufman2014,Endres2016} techniques.
The avoided-level crossing for $N$ bosons can be probed by initializing the experiment in a regime where the ground state corresponds to all the bosons occupying the right well, i.e., large values of $V_3$. 
Then, the trap depth of the right-most well is adiabatically decreased until one boson can tunnel to the left well, which first happens in the tunneling resonance.
Once the avoided level crossing is identified, the next step would be to measure the period of coherent oscillations, which can then be compared with the numerical results presented here. 

It is quite interesting to point out that the ground-state single-particle energy for the states in Figs.~\ref{fig:wavN2} and \ref{fig:wavN3} is above the classical saddle point of the external potential at $x=y=z=0$. 
Thus, the coherent oscillations found in our numerical experiments can be regarded as some form of over-the-barrier tunneling dynamics. Interestingly, it was shown in Ref.~\cite{Matveev1996} that a version of the Coulomb blockade phenomenon persists in the regime dominated by thermal activation of electrons over the tunnel barriers. 
In our case, even though the ground-state single-particle energy is above the barrier, we still observe the $\sqrt{N}$ prediction in the interaction blockade limit of the Bose-Hubbard model. This begs the question of how far above the barrier the single-particle energies can go before Eq.~\eqref{eq:bef} breaks down. In the absence of any barrier, we can consider as an example a simple harmonic oscillator potential with equidistant energies. In this case, by virtue of symmetry in the wave function, symmetric and anti-symmetric superpositions of the ground and first-excited states will still produce localized functions. However, the conceptual equivalent of tilting the potential in this case would be a simple displacement of the harmonic trap. Therefore, the single-particle energy splitting remains unchanged and no avoided crossing appears. Perhaps more important, it becomes harder to justify in the absence of a barrier the use of only the two lowest single-particle states since the energy levels are now spaced equally. These factors suggest a breakdown of Eq.~\eqref{eq:bef} in the complete absence of a barrier.
A more systematic study of this crossover behavior as the classical barrier is decreased can be an interesting direction for future work.

\begin{acknowledgments}
We thank Peter Jeszenszki and Lauri Toikka for useful discussions and Alexej Streltsov for providing access to the MCTDHB-Lab computer code. We also thank Eyal Schwartz and Pimonpan Sompet for providing relevant experimental parameters. This work has been supported by the Marsden fund of New Zealand (contract number UOO1320).
\end{acknowledgments}

\setcounter{equation}{0}
\setcounter{table}{0}

\appendix
\section{MCTDHB and Convergence}\label{appen:a}
\renewcommand{\theequation}{A\arabic{equation}}

The numerical simulations are performed using the MCTDHB method. 
Our calculations are done using the implementation in the MCTDHB-Laboratory package \cite{mctdhblab}. Here, we briefly describe the theory behind MCTDHB. We also discuss how we ascertain the numerical convergence of the MCTDHB results.

The MCTDHB method uses a variational ansatz for the many-body wave function, which must be properly symmetrized for bosons
\begin{equation}\label{mctdhbeq}
 |\Psi(t) \rangle =\sum_{n_1} \dots \sum_{n_M} C_{n_1,n_2,\dots,n_M}(t)\prod_{k=1}^M \frac{1}{\sqrt{n_k!}}[\hat{b}^{\dagger}_k(t)]^{n_k}|\mathrm{vac}\rangle,
\end{equation}
where ${\phi_k}$ is the set of single-particle functions $\phi_k(x,t)=\langle x|\hat{b}^{\dagger}_k(t)|\mathrm{vac}\rangle$. Using Lagrangian formulation, the functional action reads
\begin{align}
 S&[\{C_{\vec{n}}(t)\},\{\phi_k(x,t)\}] \\ \nonumber
 &= \int dt \biggl\{ \langle \Psi| \hat{H} - i\frac{\partial}{\partial t}|\Psi\rangle\\ \nonumber
 & - \sum_{k,j=1}^M \mu_{k,j}(t)[\langle \phi_k|\phi_j \rangle - \delta_{kj}]\biggr\},
\end{align}
where the Lagrange coefficients $\mu_{kj}$ ensure orthonormality of the single particle functions. Variations with respect to the time-varying expansion coefficients and $\{\phi_k\}$ yield a set of coupled equations of motion, which are numerically solved. Further details on the derivation can be found in \cite{AlonEtAl2008}.
The main convergence parameter in MCTDHB simulations is the number of single-particle modes $M$ used in the expansion of the many-body wave function. The importance of the single-particle functions used to expand the MCTDHB wave function can be evaluated from the eigenvalues $n^{\mathrm{NO}}_k$ of the single-particle density matrix, sometimes called as natural occupancies, $\langle \psi^{\dagger}(x)\psi(y) \rangle = \sum^M_{k=1}n^{\mathrm{NO}}_k \phi^*_k(x)\phi_k(y)$. In the subsequent discussion, we have normalized the trace of the density matrix to unity, i.e., $\sum_k n^{\mathrm{NO}}_k=1$. Furthermore, the excited eigenstates and eigenvalues shown in Fig.~\ref{fig:EN2} are obtained from the time-dependent Hamiltonian matrix calculated when performing the variational optimization of the $C_{\vec{n}}$ vectors with respect to the ground-state wave function. We have checked that the eigenenergies obtained this way are consistent with those calculated from the variationally optimized first-excited state wave functions. This procedure of obtaining the eigenspectrum using improved relaxation method has been recently applied in another work based on MCTDHB simulations \cite{Mistakidis2017}.

To exemplify how we test for numerical convergence of our results, we discuss below a specific example for $N=2$.
Relaxation towards the variationally optimized ground-state wave function is done within MCTDHB through imaginary time propagation \cite{AlonEtAl2008}.  
For our MCTDHB simulations, we have used a fast Fourier transform based grid with $\{N_x,N_y,N_z\}=\{60, 30, 30\}$ number of grid points in a box defined by $x/w_0 \in [-0.5,1.0]$, $y/w_0 \in [-0.4,0.4]$, and $z/w_0 \in [-0.4,0.4]$. We have checked that relevant properties of the low-lying energy states such as the energy splitting $\Delta E$ are converged with respect to this discretization. We present in Table~\ref{tab4} the values of the renormalized interaction couplings $\Lambda$ used in our MCTDHB simulations.
\begin{table}[!ht]
\renewcommand{\arraystretch}{2}
 \begin{tabular}{| c | c | c | c |}
 \hline \hline
  & $M=2$ & $M=3$ & $M=4$\\ \hline
 \multicolumn{1}{|c|}{$\tilde{\Lambda}$} & 0.0707128 & 0.0718848 & 0.0731065 \\ \hline \hline \hline
  & $M=5$ & $M=6$ & $M=7$ \\ \hline
 \multicolumn{1}{|c|}{$\tilde{\Lambda}$} & 0.0743990 & 0.0749100 & 0.0754298 \\ \hline \hline  
 \end{tabular}
  \caption{Renormalized interaction strength $\tilde{\Lambda}=\Lambda mw_0^2/\hbar^2$ for various $M$.
}\label{tab4}
\end{table}

\begin{figure}[!htb]
\includegraphics[width=0.48\columnwidth]{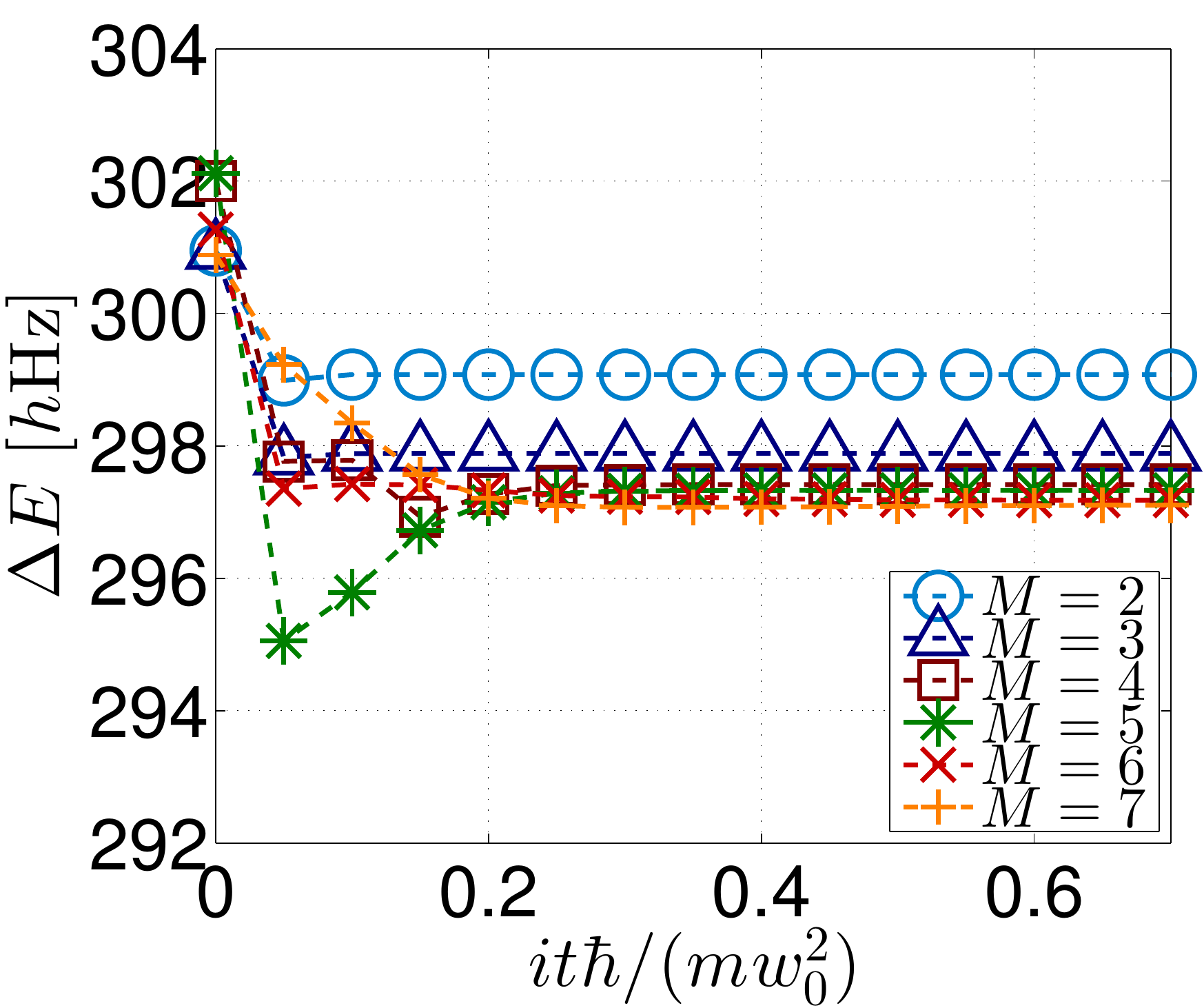}\includegraphics[width=0.5\columnwidth]{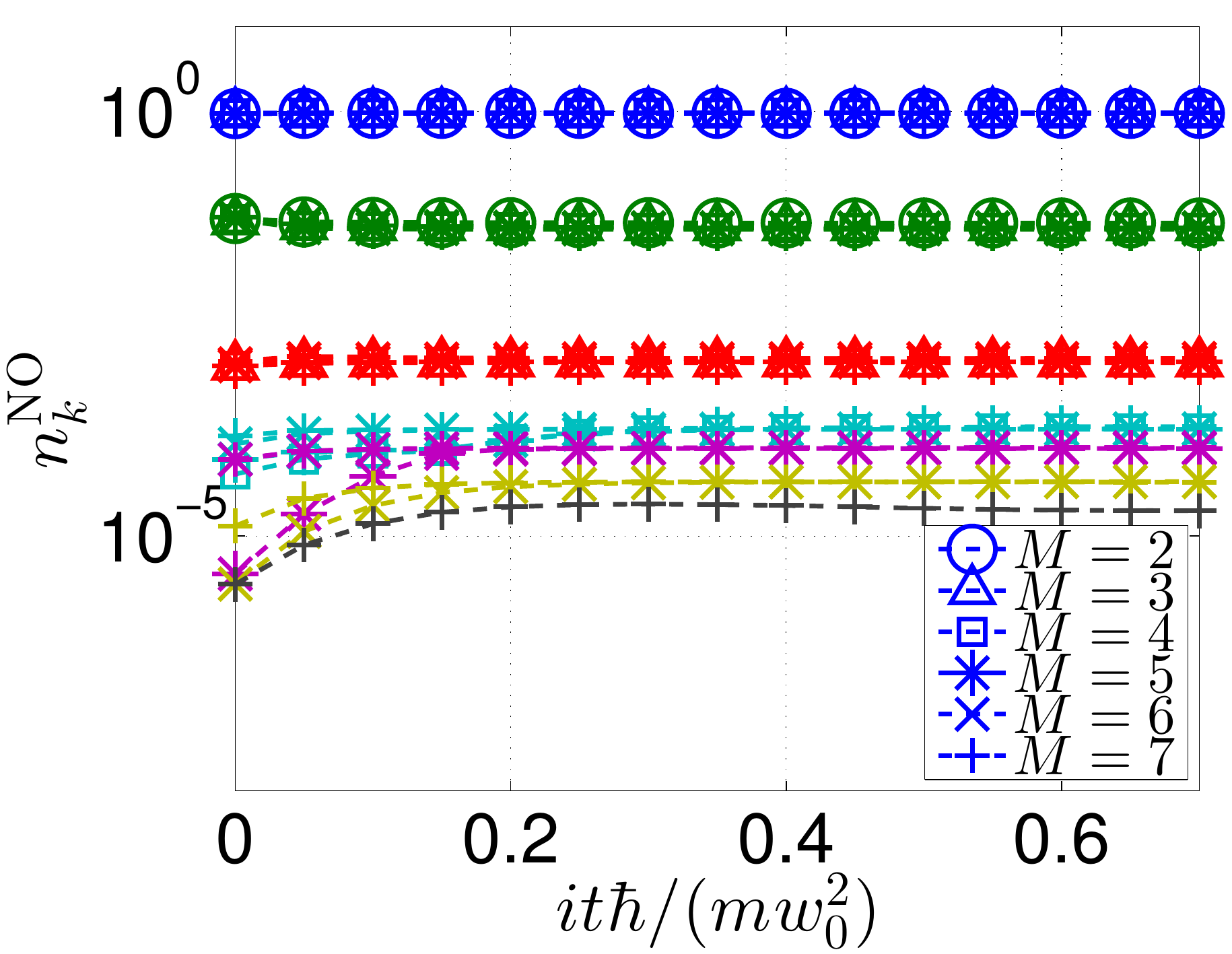}
\caption{Relaxation using imaginary time propagation within MCTDHB for $N=2$ $(^{87}\mathrm{Rb})$ at the avoided crossing. (Left) the energy difference between the ground and first-excited state. (Right) Natural occupancies of the ground-state single-partice density matrix.}
\label{fig:Eimag} 
\end{figure}

\begin{figure}[!htb]
\includegraphics[width=0.47\columnwidth]{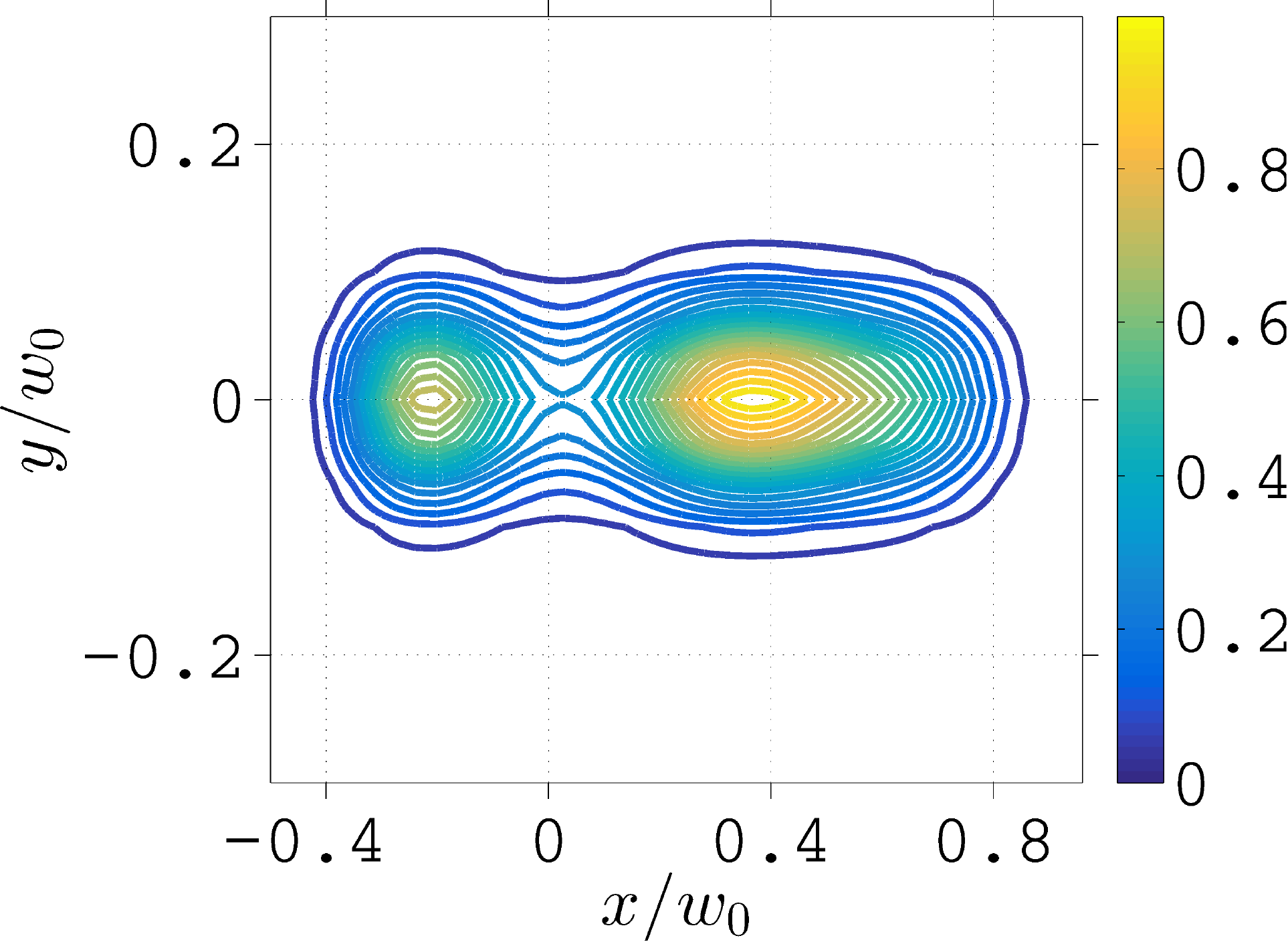}\hspace{0.2cm}\includegraphics[width=0.47\columnwidth]{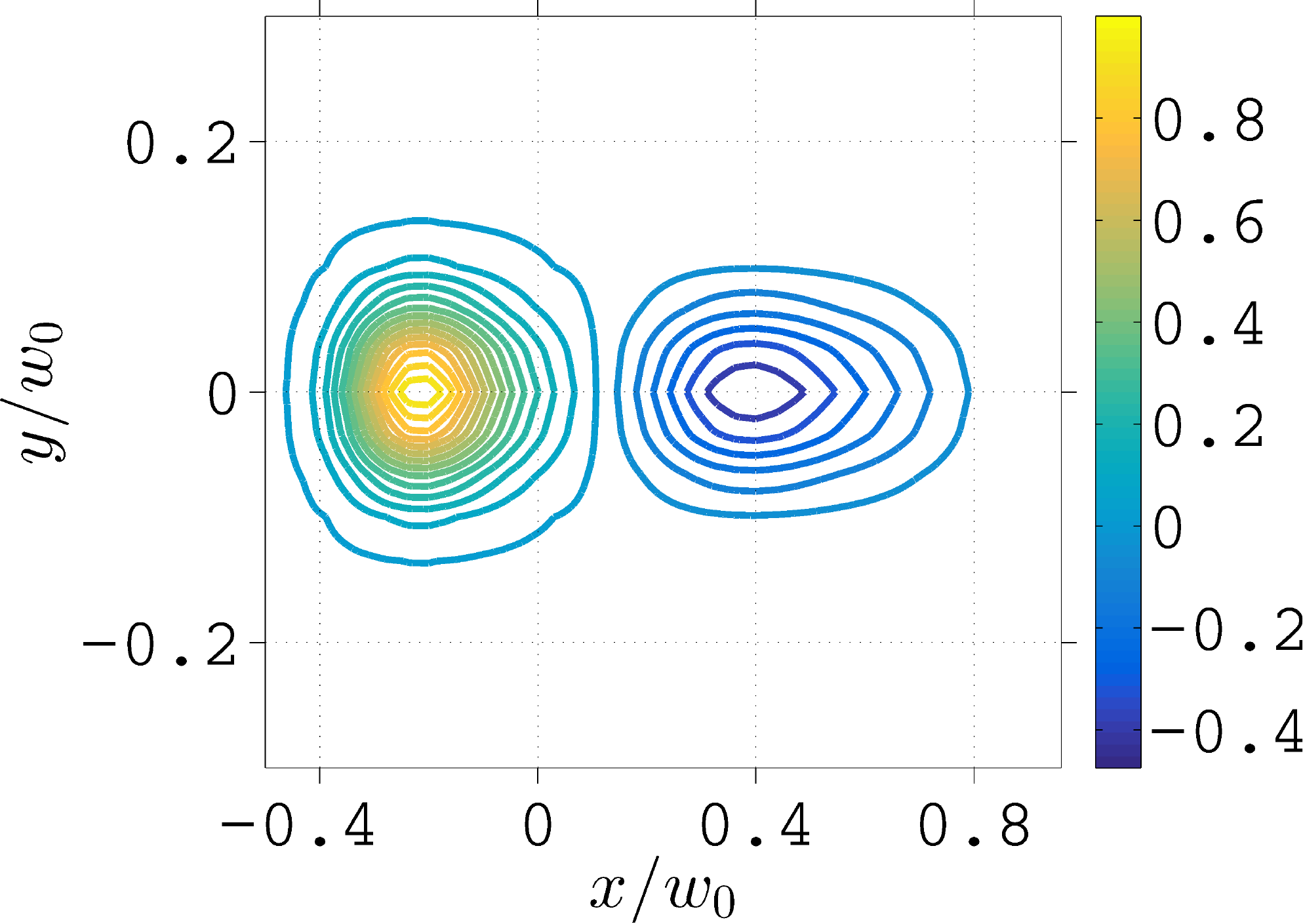}\\
\vspace{0.1cm}
\includegraphics[width=0.47\columnwidth]{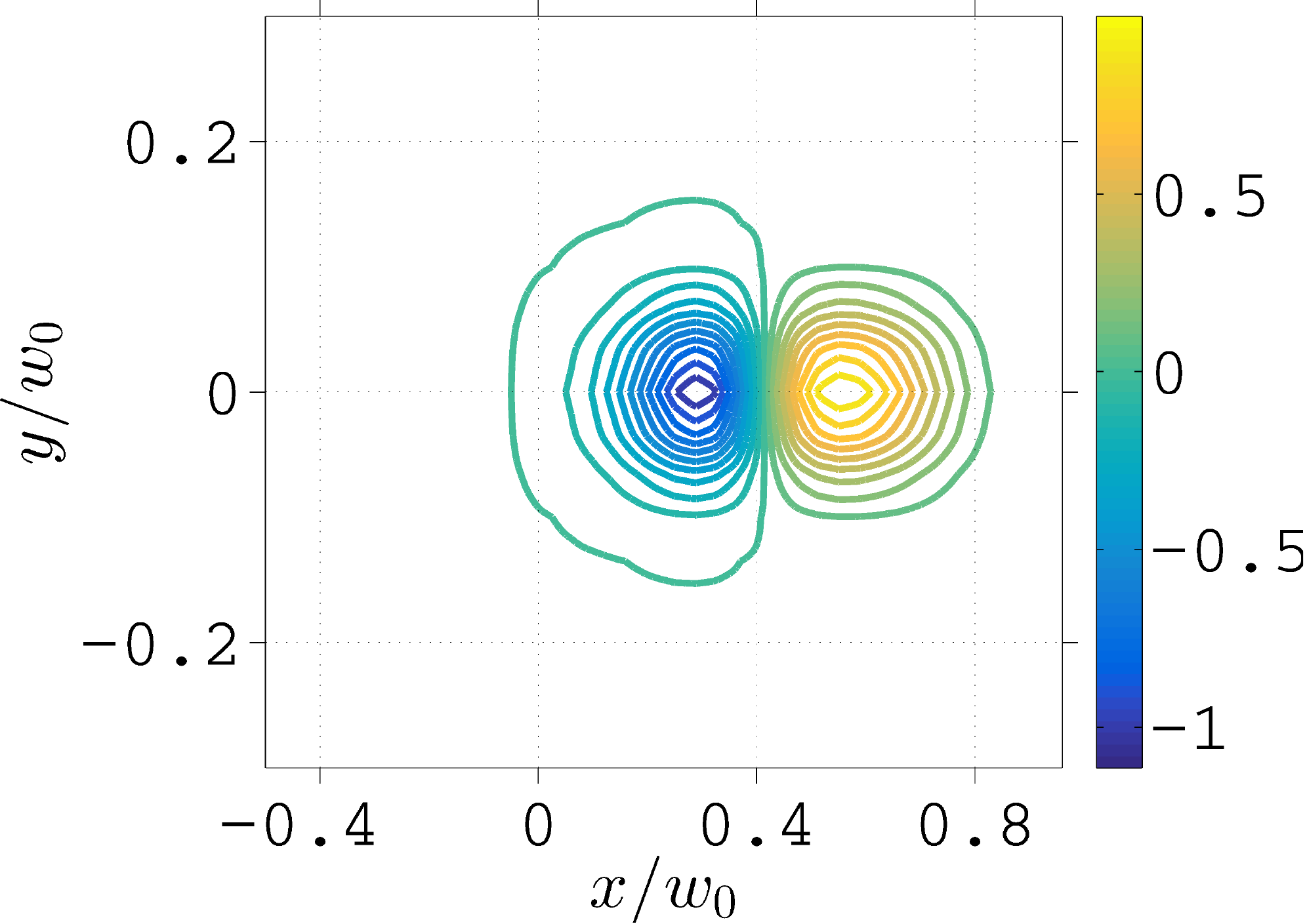}\hspace{0.2cm}\includegraphics[width=0.47\columnwidth]{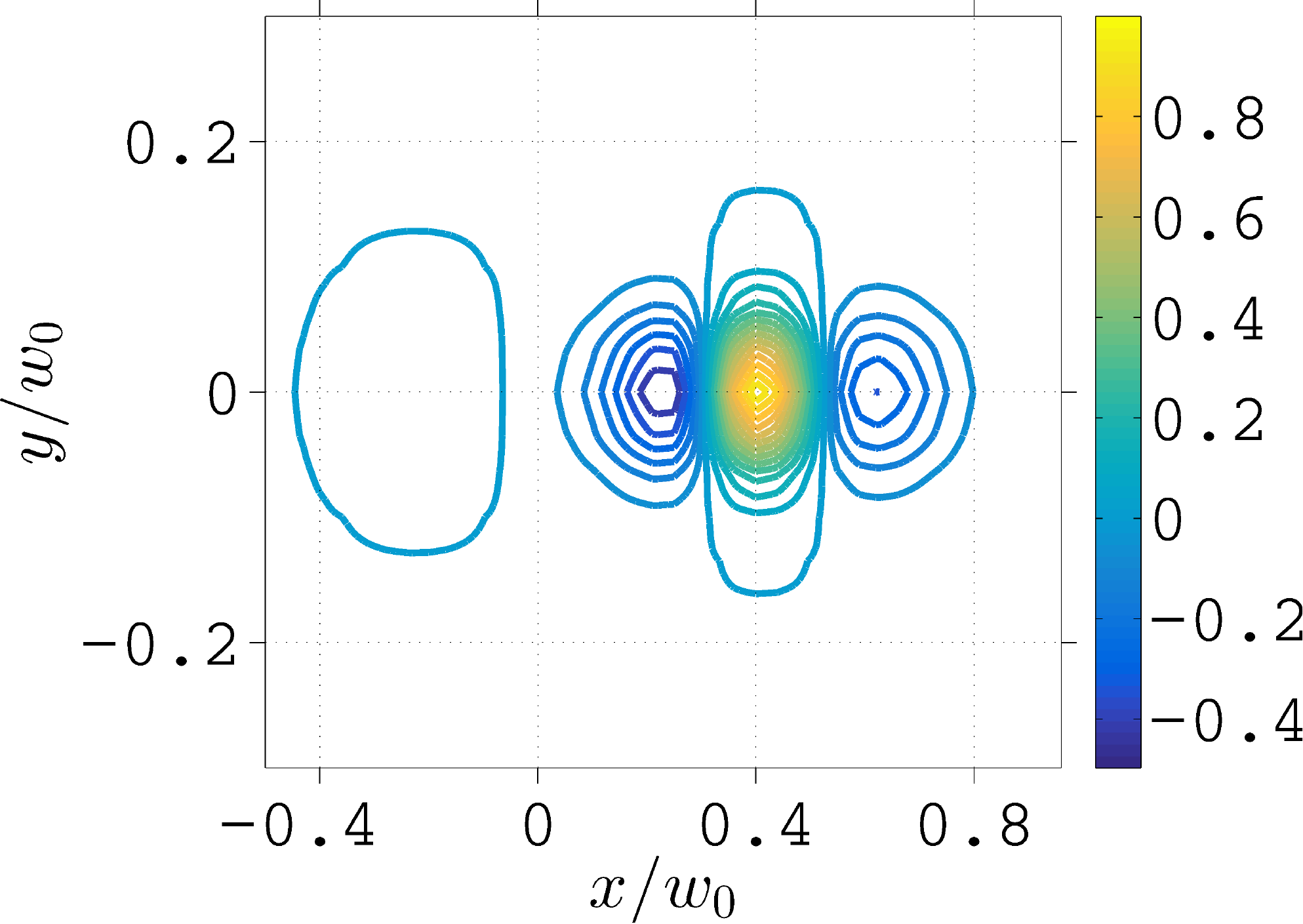}\\
\caption{Slice of the natural orbitals for the ground-state at the avoided crossing of $N=2$ and $M=4$ at $z=0$ as a function of $x$ and $y$. Top-Left: highest occupied orbital; Top-Right: second highest occupied orbital; Bottom-Right: third highest occupied orbital; Bottom-Right: least occupied orbital.}
\label{fig:natorb} 
\end{figure}

Since we are interested in the energy splitting between the two lowest energy states, we show in Fig.~\ref{fig:Eimag} the energy difference $\Delta E$ as a function of imaginary time for different single-particle modes $M$. It can be seen that already for $M=4$ and $M=5$, the energy splitting has nicely converged. Moreover, this also confirms that the renormalization procedure described above has worked. The time-evolution of the natural occupancies of the ground-state single-particle density matrix is also shown in Fig.~\ref{fig:Eimag}. The convergence towards the relaxed energy splitting coincides with an increase in the fourth lowest natural occupancy. This observation reveals the importance of using at least $M=4$ modes in MCTDHB to simulate the stationary states of the system. It is insightful to look at the natural orbitals for $M=4$ shown in Fig.~\ref{fig:natorb}. The symmetry of the external potential along the $y$-direction is trivially reflected in the symmetric nature of the orbitals about the $y$-axis. The $x$-direction or the axis of asymmetry is more interesting since here we can see from the nature of the two lowest occupied natural orbitals that MCTDHB needs at least $M=4$ modes in order to capture the finer details of the external potential brought about by the fact that two optical tweezers actually form the wider well. This is apparent from the presence of nodal structures in the region occupied by the wide well as depicted in the plots in the bottom panel of Fig.~\ref{fig:natorb}.
Multi-mode generalizations of a bosonic Josephson junction \cite{Smerzi1997,Milburn1997,Albiez2005,Levy2007} have been explored in various works demonstrating the importance of going beyond the two-mode approximation provided by the two lowest single-particle modes in order to predict novel quantum phenomena \cite{Massiello2005,Dounas2007,Zollner2008,Sakmann2009,Sakmann2010,Garcia2012,Cosme2015,Brouzos2015}.

Recently, pathologically slow convergence of MCTDHB results was demonstrated for nonequilibrium dynamics of attractively interacting bosons in the absence of an external trap \cite{Cosme2016}.
However, for the system considered in this work, convergence from using a relatively fewer number of single-particle modes $M$ is to be expected due to tight three-dimensional trapping geometry leading to a large separation of the two lowest single-particle energies from the rest of the spectrum as seen from Fig.~\ref{fig:spen}. Moreover, the bosons are repulsively interacting and the external trap is always present in our simulations. Good agreement with the Bose-Hubbard results for $N=2$ as shown in Fig.~\ref{fig:EN2} further validates the applicability of MCTDHB for the type of problem considered here.

\section{Effective Bose-Hubbard parameters}\label{appen:b}
\renewcommand{\theequation}{B\arabic{equation}}

In this section, we briefly discuss the recipe for calculating the effective Bose-Hubbard parameters according to Ref.~\cite{Wall2015}.
The first step is to obtain a set of $\tilde{M}$ single-particle eigenfunctions  $\{\psi_1,\psi_2,\dots,\psi_{\tilde{M}}\}$ for a specific realization of the 3D potential.
Specifically for our simulations, we are using the sinc-DVR basis \cite{Wall2015} with the same grid size and box size as described in Appendix~\ref{appen:a}.
We seek linear combination of these states to write the localized or ``Wannier'' function as 
\begin{equation}\label{eq:wann}
  \phi_{\ell}(\mathbf{r}) \equiv \phi(\mathbf{c},\mathbf{r}) = \sum_{i=1}^{\tilde{M}} c^{\ell}_i \psi_i(\mathbf{r}),
\end{equation}
where $\mathbf{c} \cdot \mathbf{c} = 1$. 
Following the prescription for construction of localized Wannier functions in \cite{Wall2015}, we calculate the localization functional
\begin{equation}
 \mathcal{L}(\psi_i,\psi_j) = \int dy~dz \int_{-\infty}^{x_{\mathrm{cut}}} dx \psi^{*}_i(\mathbf{r})\psi_j(\mathbf{r}),
\end{equation}
where $x_{\mathrm{cut}}$ defines the left well say the position of the local maxima at $x=0$. Both $\mathbf{c}$ and the set of single-particle states must satify the condition
\begin{equation}
 \mathrm{max}~\mathcal{L}[\phi(\mathbf{c},\mathbf{r}),\phi(\mathbf{c},\mathbf{r})].
\end{equation}
Alternatively, this condition can be written as
\begin{equation}
 \mathrm{max}~\mathbf{c} \cdot \mathbb{L} \cdot \mathbf{c},
\end{equation}
where we define the localization matrix as
\begin{equation}
 \mathbb{L}_{ij} = \mathcal{L}[\psi_i(\mathbf{r}),\psi_j(\mathbf{r})].
\end{equation}
Thus, the vector $\{c^{\ell}_i\}$ in Eq.~\eqref{eq:wann}, which maximizes the localization in the $\ell$-site, is one of the $\tilde{M}$ eigenvectors of the localization matrix. 
In this work, we truncate this expansion and use only the two lowest single-particle eigenstates, i.e., $\tilde{M}=2$. For a symmetric double-well potential, we recover for the $\mathbf{c}$ vectors the symmetric and anti-symmetric superposition of the two lowest-lying single-particle energy states, $(1,1)^{\mathrm{T}}/\sqrt{2}$ and $(1,-1)^{\mathrm{T}}/\sqrt{2}$, respectively.
This procedure allows us to construct the two localized Wannier functions $\phi_{\ell}$ for arbitrary choice of $V_3$, we can then obtain the effective Bose-Hubbard parameters using the two lowest single-particle modes. The tunneling term is obtained from
\begin{equation}
 J = \int d\mathbf{r} \phi^{*}_{\ell}(\mathbf{r})\biggl[ \frac{-\hbar^2}{2m}\nabla^2 + V(\mathbf{r}) \biggr]\phi_{\ell'}(\mathbf{r}),
\end{equation}
where $\ell \neq \ell'$.
The on-site energies are given by
\begin{equation}
 \epsilon_{\ell} = \int d\mathbf{r} \phi^{*}_{\ell}(\mathbf{r})\biggl[ \frac{-\hbar^2}{2m}\nabla^2 + V(\mathbf{r}) \biggr]\phi_{\ell}(\mathbf{r}).
\end{equation}
The on-site interactions are given by
\begin{equation}
 U_\ell = \frac{4\pi\hbar^2a_s}{m}\int d\mathbf{r} |\phi_{\ell}(\mathbf{r})|^4.
\end{equation}
where $a_s$ is the $s$-wave scattering length and for our simulations we use $a_s=5.45~\mathrm{nm}$. 

\section{Fidelity calculation between MCTDHB wave functions}\label{appen:c}
\renewcommand{\theequation}{C\arabic{equation}}
In order to make sure that the suggested protocol in measuring the enhancement factor is sensible, we have suggested to calculate the fidelity
\begin{equation}
 \mathcal{F}=|\langle \Psi|\chi \rangle|^2.
\end{equation}
For brevity, we will drop the time-dependence on the wave functions since we are only interested in overlaps between relaxed or stationary states. Moreover, we shall assume that both MCTDHB wave functions use the same number of single-particle modes $M$. In this case, the MCTDHB wave functions can be written as 
\begin{align}
 |\Psi\rangle &= \sum_n C_n | n\rangle \\ \nonumber
 |\chi\rangle &= \sum_m B_m | m\rangle,
\end{align}
where
\begin{equation}
 |n\rangle = \prod_{k=1}^M \frac{[\hat{b}^{\dagger}_k]^{n_k}}{\sqrt{n_k!}}|0\rangle
\end{equation}
and
\begin{equation}
 |m\rangle = \prod_{k=1}^M \frac{[\hat{d}^{\dagger}_k]^{m_k}}{\sqrt{m_k!}}|0\rangle.
\end{equation}
The bosonic creation operators create a boson in a single-particle function according to
\begin{align}\label{eq:bosac}
 \hat{b}^{\dagger}_k | \mathrm{vac} \rangle &= |\phi_k \rangle \\ \nonumber
 \hat{d}^{\dagger}_k | \mathrm{vac} \rangle &= |\varphi_k \rangle.
\end{align}
Obtaining the fidelity equates to calculation of the overlap
\begin{equation}
 \langle \Psi|\chi \rangle = \sum_{m,n} C^*_n B_m \langle n|m \rangle
\end{equation}
In general, we can easily access the variationally optimized expansion coefficients $C_n$ and $B_m$. However, obtaining the overlaps $\langle n|m \rangle$ is not trivial because the basis functions between two different MCTDHB wave functions are in most cases not orthonormal, i.e., $\langle \phi_i| \varphi_k \rangle \neq \delta_{i,k}$. 
In order to proceed further, we expand both of the single-particle functions $|\phi_i\rangle$ for $|n\rangle$ and $|\varphi_i\rangle$ for $|m\rangle$ in terms of a primitive basis $\{\zeta_j\}$ defined by the chosen grid discretization scheme. Note that we are now assuming that the two wave functions are solved in the same underlying primitive basis. Then, we can write
\begin{equation}
 |\phi_i\rangle = \sum f_{ij} | \zeta_j \rangle
\end{equation}
and
\begin{equation}
 |\varphi_i\rangle = \sum g_{ij} | \zeta_j \rangle.
\end{equation}
Since each of the single-particle basis functions form an orthonormal set, we can then calculate the overlap matrix that can be used to expand one single-particle function in terms of the other
\begin{align}
 |\varphi_i\rangle &= \sum_{k} |\phi_k\rangle \langle \phi_k| \varphi_i\rangle = \sum_{k,\ell} f^*_{k\ell}g_{i\ell}|\phi_k\rangle \\ \nonumber
 &=\sum_k A_{ki} |\phi_k\rangle,
\end{align}
where we define the overlap matrix as
\begin{equation}\label{eq:ovA}
 A_{ki} = \sum_{\ell} f^*_{k\ell}g_{i\ell}.
\end{equation}
Using Eqs.~\eqref{eq:bosac} and \eqref{eq:ovA}, it is straightforward to obtain the corresponding commutation relation
\begin{equation}\label{eq:ajk}
 [\hat{b}_j,\hat{d}^{\dagger}_k] = A_{jk}.
\end{equation}
This commutation relation allows us to calculate overlaps for arbitrary number of modes $M$ and number of bosons $N$. For $N=2$, it can be shown that
\begin{align}
 \langle n|m \rangle &= \frac{1}{\mathcal{N}} \langle \mathrm{vac}| \hat{b}_i \hat{b}_j \hat{d}^{\dagger}_k \hat{d}^{\dagger}_\ell|\mathrm{vac} \rangle \\ \nonumber
 &=\frac{1}{\mathcal{N}}\left(A_{ik}A_{j\ell}+A_{jk}A_{i\ell} \right),
\end{align}
where $\mathcal{N}=\sqrt{n_i!n_j!m_k!m_\ell!}$.
For $N=3$, the associated overlap is 
\begin{align}
 \langle n|m \rangle &= \frac{1}{\mathcal{N}} \langle \mathrm{vac}| \hat{b}_i \hat{b}_j \hat{b}_s \hat{d}^{\dagger}_t \hat{d}^{\dagger}_k \hat{d}^{\dagger}_\ell|\mathrm{vac} \rangle \\ \nonumber
 &=\frac{1}{\mathcal{N}}[A_{it}(A_{jk}A_{s\ell}+A_{sk}A_{j\ell}) \\ \nonumber
 &+ A_{jt}(A_{ik}A_{s\ell}+A_{sk}A_{i\ell}) + A_{st}(A_{ik}A_{j\ell}+A_{jk}A_{i\ell})],
\end{align}
where now $\mathcal{N}=\sqrt{n_i!n_j!n_s!m_t!m_k!m_\ell!}$.

\bibliography{biblio}

\end{document}